\documentclass[12pt]{article}
\usepackage[top=3cm, bottom=3cm, left=3cm, right=3cm]{geometry} 
\usepackage{color}
\usepackage{rotating}
\usepackage{graphicx}
\usepackage{amsmath}
\usepackage{amssymb}
\usepackage{float}
\usepackage{subfig}
\usepackage[countmax]{subfloat}
\usepackage{hyperref}
\def\displayandname#1{\rlap{$\displaystyle\csname #1\endcsname$}%
                      \qquad \texttt{\char92 #1}}

\begin{document}
\title{\bf{Simulation analysis with rock muons from atmospheric neutrino interactions in the ICAL detector at INO}} 

\author{R. Kanishka\footnote{
Email: kanishka.rawat.phy@gmail.com}~$^{, \#}$, D. Indumathi$^{\dagger,\oplus}$, Lakshmi S. Mohan$^{\circ}$, V. Bhatnagar$^{\ddagger}$
\\
  \\
  {\it $^{\#}$Department of Physics, Chandigarh University, Gharuan, Mohali, 140413, India}, \\
  {\it $^\dagger$The Institute of Mathematical Sciences, Chennai
    600113, India}, \\
  {\it $^\oplus$Homi Bhabha National Institute, Mumbai 400094, India.}\\
  {\it $^{\circ}$National Centre for Nuclear Research, 7 Pasteura str., 02-093, Warsaw. }\\
{\it $^\ddagger$Physics Department, Panjab University, Chandigarh
160014, India} \\
}

\maketitle

\begin{abstract}

The proposed magnetized Iron CALorimeter detector (ICAL) to be built
in the India-based Neutrino Observatory (INO) laboratory aims to study
atmospheric neutrinos and its properties such as precision measurements
of oscillation parameters and the neutrino mass hierarchy. High energy
charged current (CC) interactions of atmospheric neutrinos with the rock
surrounding the detector produce so-called ``rock muons'' along with
hadrons. While the hadron component of these events are absorbed in the
rock itself, the rock muons traverse the rock and are detected in the
detector. These rock muon events can be distinguished from cosmic muons
only in the upward direction and can provide an independent measurement
of the oscillation parameters. A simulation study of these events at the
ICAL detector shows that, although reduced in significance compared to
muons produced in direct CC neutrino interactions with the detector, these
events are indeed sensitive to the oscillation parameters, achieving a
possible $1\sigma$ precision of 10\% and 27\% in determining $\Delta
m_{32}^2$ and $\sin^2\theta_{23}$, respectively. Hence a combination of
the standard atmospheric neutrino analysis which is the main goal of
ICAL, with these rock muon events, will improve the precision
reach of ICAL for these parameters.

\end{abstract}

\newpage

\section{Introduction}
\label{intro}

Neutrinos as described by the ``Standard Model''
are massless and occur in three distinct flavours
$\nu_{e}$, $\nu_{\mu}$, $\nu_{\tau}$. Many experiments on solar
\cite{fukuda,fukuda1,ahmad,aharmim,smy,ahmedsn,yoo,cleveland,hampel,abdurashitov,altmann}, atmospheric \cite{fukuda0,allison,ambrosio,ambrosio1,sanchez},
accelerator~\cite{ahnmh}, and reactor~\cite{eguchi} neutrinos have
confirmed that neutrinos have mass, and the flavour states are mixtures
of mass eigenstates ($\nu_{1}$, $\nu_{2}$, $\nu_{3}$) with different
masses---this leads to the phenomenon of neutrino oscillations and is an
important piece of evidence for physics beyond the Standard Model. The
PMNS neutrino mixing matrix~\cite{pontecorvo,maki}, can be parametrized in
terms of three mixing angles and a charge conjugation-parity violating
(CP) phase $\delta_{CP}$. Some recent results from reactor neutrino
experiments ~\cite{reno,daya} have reconfirmed the oscillations in the
neutrino flavours and non-zero value of the across-generation mixing
angle $\theta_{13}$. Apart from this, these experiments determine the
extent of mixing and the differences between the squared masses as well,
although not the absolute values of the masses themselves.

There are three possible arrangements of the neutrino masses. For normal
ordering, we have\footnote{We have used natural units with $\hslash=c=1$.}
$m_{1} < m_{2} \ll m_{3}$; hence, $\Delta m^{2}_{32} \equiv m_{3}^{2} -
m_{2}^{2} > 0$ $eV^2$ and $m_{3}$ $\gtrsim \surd \Delta m^{2}_{32} \simeq
0.03-0.07$ eV.  For inverted ordering, $m_{2} \gtrsim m_{1} \gg m_{3}$
with $m_{1,2} \gtrsim \surd \Delta m^{2}_{23}
\simeq 0.03-0.07$ eV;
hence, $\Delta m^{2}_{32} \equiv m_{3}^{2} - m_{2}^{2} < 0$ $eV^2$. In
the degenerate case, $m_{1} \simeq m_{2} \simeq m_{3}$. If the ordering
is strong, then the ordering also determines the mass hierarchy; this
problem~\cite{physics1} is still not solved and it is believed that
upcoming experiments like the magnetized Iron Calorimeter detector (ICAL)
at the proposed India-based Neutrino Observatory (INO)~\cite{ICAL:2015stm}
will be able to answer this problem along with the precision of oscillation parameters. Note that the sign of $\Delta m^{2}_{32}$
(or, equivalently, the sign of $\Delta m^{2}_{31}$) which determines the
mass ordering is still unknown, as well as the octant of $\theta_{23}$.
Precision experiments sensitive to matter effects during propagation
of the neutrinos through the Earth can determine the sign of $\Delta
m^{2}_{31}$. Another open question regarding the neutrinos is, whether
there is CP violation in the leptonic sector. Experiments such as DUNE
\cite{Dune} and JUNO \cite{Juno} will also be sensitive to the currently unknown
oscillation parameters such as the mass hierarchy and the CP phase.

The ICAL detector at the proposed INO lab will be a 51 kton magnetized
iron detector with layers of iron of 56 mm thickness interspersed with
active resistive plate chamber (RPC) detectors in the 40 mm air gap. It
will be optimized to study muons produced in the charged-current (CC)
interactions of atmospheric neutrinos with the detector. Hence it is
also suitable for measuring the so-called upward-going muon flux due to
CC interactions of the atmospheric neutrinos with the rock surrounding
the detector (the downward-going muon flux is swamped by the cosmic ray
muon background and is therefore not useful for neutrino oscillation
studies)~\cite{upmuon3}. The muon loses an unknown fraction of its energy
while traversing the rock to reach the detector; it still carries an
imprint of the oscillations of the parent neutrino that produced it. It
is therefore useful to study these upward-going or rock muons that have a
characteristic signature in the detector. In this paper, we discuss the
simulation studies in the ICAL detector at INO using upward-going muons
and their significance. This study allows us an independent measurement
of oscillation parameters.

The paper is organized as follows. In section~\ref{up}, we briefly discuss the upward-going muons at the ICAL detector. In section~\ref{det_res}, we discuss the detector response for such muons. In section~\ref{back} we discuss the main backgrounds to the rock muon events. In section~\ref{phy_an} the detailed simulation procedure including data generation, methodology for oscillation studies and $\chi^{2}$ analysis are described. In section~\ref{osc_res}, we discuss the results: sensitivity to measurements of the oscillation parameters, and comparison of ICAL sensitivity with existing data \cite{nitta}. We conclude with discussions in section~\ref{diss_con}.

\section{Upward-going Muons at the ICAL Detector} \label{up}

The ICAL detector will be located under 1279 m (approx) high mountain peak and will have a minimum rock cover of about 1 km in all directions, thereby reducing the cosmic muon backgrounds \cite{cosmic}. Upward-going muons arise from the interactions of atmospheric neutrinos with the rock material surrounding the detector, typically within the range of $\sim$200 m (beyond this distance, the muon energy loss is so large that only very high energy muons can reach the detector, but the flux of such events is very small). These upward-going muons, also known as rock muons \cite{upmuon,upmuon1,upmuon2}, provides an independent measurement of the oscillation parameters, although the sensitivity of upward-going muons to the oscillation parameters is lower than contained-vertex muons produced by atmospheric $\nu_\mu$ interactions inside ICAL. But an independent measurement using upward-going muon provides a consistency check with the contained vertex analysis, that results in a slight improvement of the overall measurement. This kind of analysis is helpful in any neutrino experiment.

\begin{figure}[bth]
\centering
\includegraphics[width=0.48\textwidth]{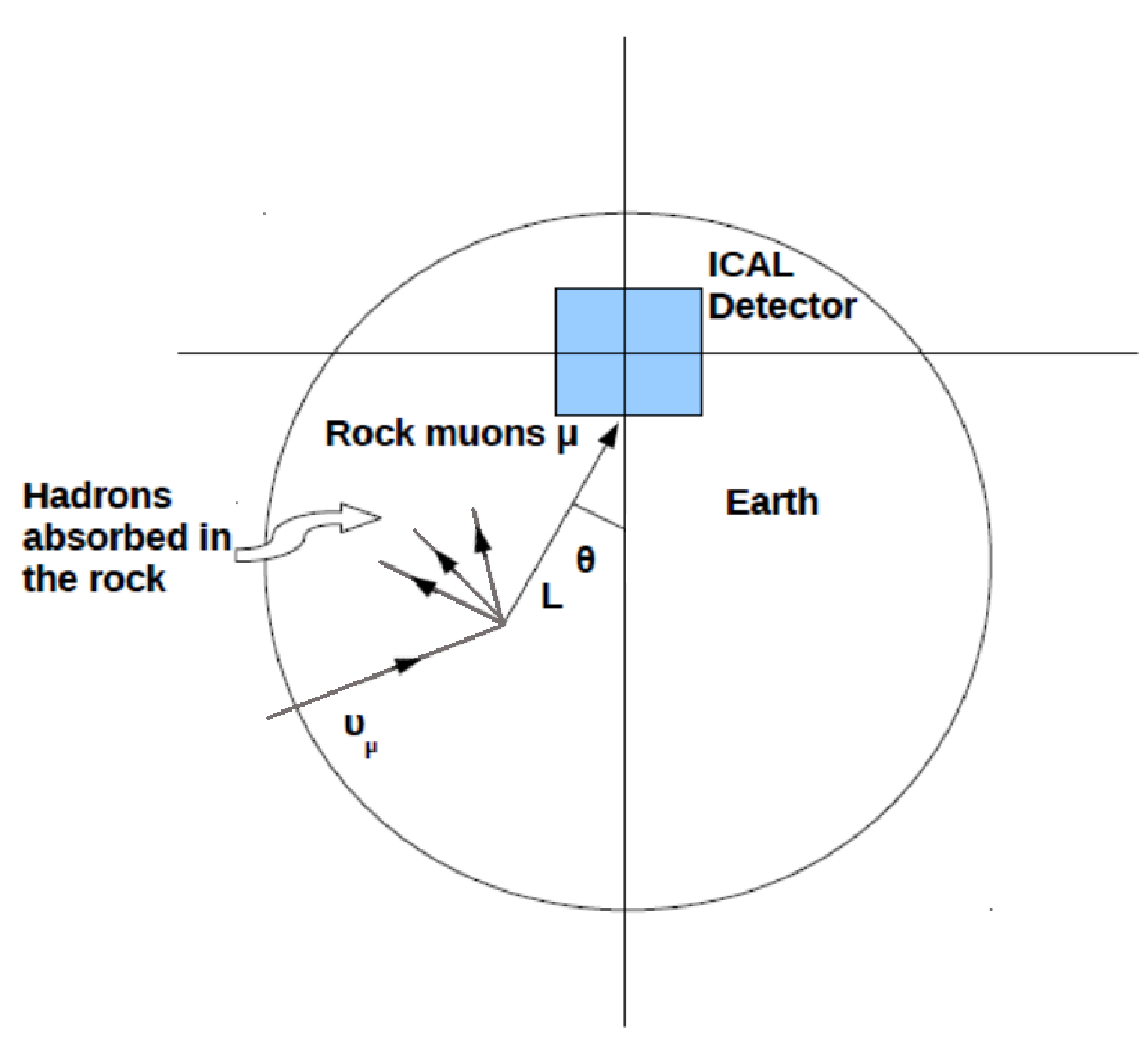}
\caption{Schematic (not to scale) of the processes that give rise to
upward-going muons at ICAL.}
\label{fig:rock-mu}
\end{figure}

Figure~\ref{fig:rock-mu} shows a schematic of the processes that give rise
to upward-going muons at the ICAL detector. Neutrinos, after interacting
with rock, produce hadrons and muons. The hadrons get absorbed in the rock
and the upward-going muons travel a distance $L$ making an angle $\theta$
with the $z$-axis, and finally reach the ICAL detector. Due to the kinematics,
especially at higher neutrino energies, upward-going neutrinos also
(dominantly) produce upward-going muons.

The muon loses a substantial part of its energy (on the average) in the
rock before it reaches the detector so the oscillation signature becomes
more complicated. The formula for the average muon energy loss for muons
of energy $E_{\mu}$~\cite{Workman:2022ynf} produced in the rock is given by,
\begin{eqnarray}
 \frac{d E_{\rm \mu}} {d x} & = & -a -b E_{\rm \mu}~,
  \label{mu_loss1}
\end{eqnarray}
so the energy loss of the muons after propagation through a distance $X$
g/cm$^{2}$ is:
\begin{eqnarray}
  E_{\rm \mu} = (E^{0}_{\rm \mu} + \epsilon) \exp(-b X)
- \epsilon~,
\label{mu-loss2}
\end{eqnarray}
where $\epsilon = a/b$, $E^{0}_{\rm \mu}$ is the initial muon energy,
$a$ accounts for ionization losses and $b$ accounts for the three
radiation processes: bremsstrahlung, photoproduction and production of
electron-positron pairs. We have $\epsilon$ = $a/b$ $\sim$ 500 GeV, where
both $a$ and $b$ depends on $E_{\rm \mu}$. This formula is approximate
and indicative of the kind of energy loss at different energies. The
actual upward-going muons have been simulated using the NUANCE neutrino
generator~\cite{nuance} which takes into consideration the energy
dependence of $a$ and $b$ as well as accounts for fluctuations. More
details on the generator are given below; here we only mention that the
NUANCE generator lists the vertex and energy-momentum of all the final
state particles produced in CC interactions of muon neutrinos with the
rock material. Hence it is possible to use the energy and direction
information of the muon at the production point along with the energy
loss formulae in Eqs.~\eqref{mu_loss1} and \eqref{mu-loss2} to propagate
the muon to the closest surface of the detector.

The muon energy loss depends on the distance $L$ traversed through the
rock of density $\rho(L)$ from the production point (vertex of the CC
interaction of the muon neutrino with rock) to the detector, which
we have taken to be $\sim$ 1 km underground. Since the muons travel
essentially in the Earth's crust (which is about 30--40 km thick on
average), hence we have $X = \rho \, L$; with the rock density $\rho = 2.65$
gm/cc, as per geo-technical studies carried out by the collaboration
\cite{ICAL:2015stm}.

\begin{figure}[bth]
\centering
\includegraphics[width=0.49\textwidth]{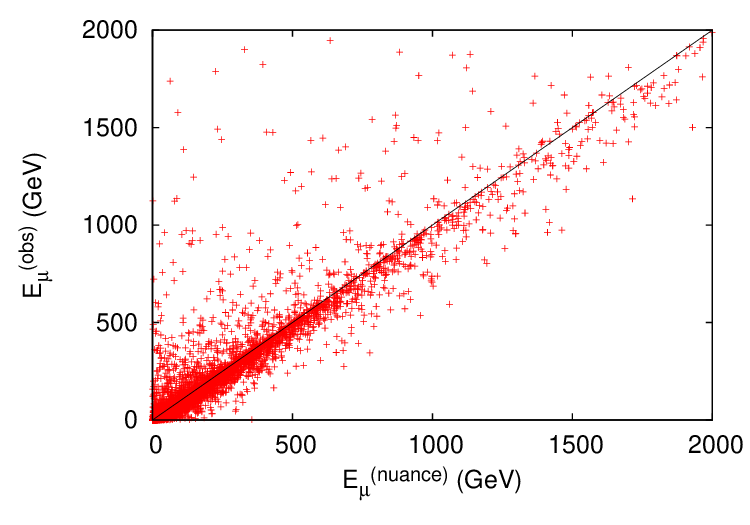}
\includegraphics[width=0.47\textwidth]{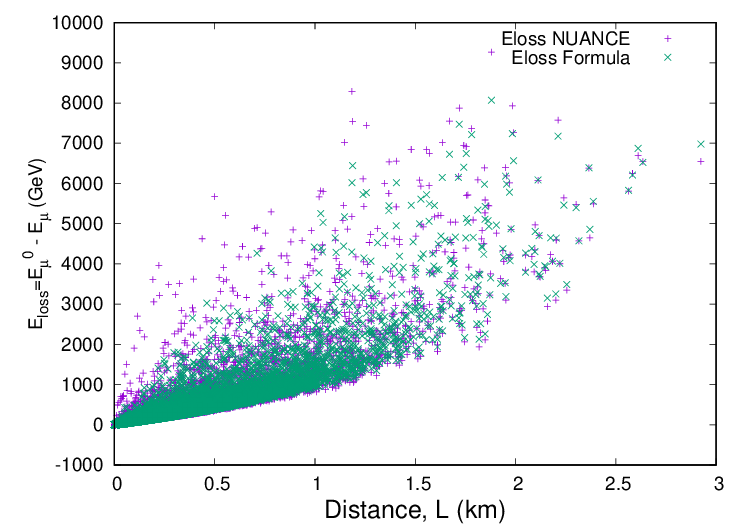}
\caption{Comparison of the energy $E_\mu^{(cal)}$ of upward-going muons 
calculated from Eq.~\eqref{mu-loss2} (black line) vs. $E_\mu^{(nuance)}$
calculated by the NUANCE neutrino generator including fluctuations
(scatter plot) (top panel). Energy loss of the muon (energy at production
minus the energy at the detector) as a function of the path length
travelled in the rock. (bottom panel)}
\label{fig:scatter}
\end{figure}

Figure~\ref{fig:scatter} (top panel) shows the energy $E_\mu^{(cal)}$ of
muons arriving at the detector from all possible upward directions as
calculated from Eq.~\eqref{mu-loss2} using fixed values of $a (= 2.68$
MeV cm$^2$/g) and $b (= 3.92 \times 10^{-6}$ cm$^2$/g), versus the energy
$E_\mu^{(nuance)}$ obtained from NUANCE for the total generated sample
of 200 years exposure at ICAL. The figure shows the relative energy
composition of the rock muons that reach the detector. The dependence is
linear on the average, but the NUANCE points show large fluctuations,
especially at lower energies which are of interest for atmospheric
neutrino studies.

Figure~\ref{fig:scatter} (bottom panel) shows the
relative energy loss of the muon as a function of the distance
travelled in the rock. It can be seen that only higher energy muons can
reach the detector from further away, as expected. Each point is
generated by (a) using the NUANCE value for the energy at the
production point and at the detector (purple points), and (b) by using the NUANCE value
at production and the energy loss formula in Eqs.~\eqref{mu_loss1} and
\ref{mu-loss2} to determine the muon energy at the detector (green
points). The difference between the two sets of points is due to
the correct use of the energy-dependent values of $a$ and $b$ and also
the inclusion of fluctuations, which makes a large difference at lower
energies for the NUANCE points.

A substantial number of muons are absorbed before they reach the
detector. Though the observed number is small, they carry important
signatures of neutrino oscillations. The muon neutrino survival probability $P_{\mu\mu}$
(which goes to 1 as $E_\nu$ increases) in vacuum is given by,
\begin{eqnarray}
P_{\mu\mu} = 1 - \sin^{2}2\theta_{23} \sin^{2}
		\frac{1.27 \Delta m^{2}_{32} L_{\nu}}{E_{\nu}}~,
\label{eq1}
\end{eqnarray}
where the neutrino path length, energy and mass squared
differences, $L_{\nu}$, $E_{\nu}$, $\Delta m^{2}_{ij} \equiv
m_i^2 - m_j^2$, are in units of km, GeV and eV$^{2}$ respectively. The
muon neutrino survival probability $P^m_{\mu\mu}$ in matter can be
approximated by~\cite{octant1,Indumathi:2004kd}:
\begin{eqnarray} \nonumber
P^m_{\mu\mu} & \approx & 1 - \sin^2 2\theta_{13}^{m} \sin^4\theta_{23}
\sin^2\Delta m^2_{31,m}~ - \sin^2 2\theta_{23} \times \nonumber \\
   & & \left[\sin^2\theta_{13}^{m} \sin^2\Delta m^2_{21,m} +
    \cos^2\theta_{13}^{m} \sin^2\Delta m^2_{32,m} \right]~,  \\
  & \equiv & P^{(2)}_{\mu\mu} - \sin^2\theta_{13} \times
  \left[\frac{{A}}{{\Delta-A}} T_1 + \right. \nonumber \\
     & & \left. \left(\frac{\Delta}{{\Delta-A}}\right)^2
    \left(T_2 \sin^2[({{\Delta -A}}) x] +
    T_3\right) \right]~.
\label{eqP}
\end{eqnarray}
Here $P^{(2)}_{\mu\mu}$ is the (2-flavour) matter-independent survival
probability, $x \equiv 1.27 L (\hbox{km})/E (\hbox{GeV})$ in units of
eV$^{-2}$  and the matter term is given by $A = \pm 7.6 \times 10^{-5} \rho$
(gm/cm$^3)E$ (GeV), also in the same units (where the $\pm$ signs apply to
neutrinos and anti-neutrinos respectively). Here
$\Delta m^2_{ij,m}$ are dimensionless quantities related to the mass
squared differences in matter; see Ref.~\cite{octant1} for more details. We
have denoted the dependence on the dominant mass squared difference as
$\Delta$: $\Delta \sim \Delta m^2_{32} \sim \Delta m^2_{31}$, and in
the last line $T_{1,2,3}$ are coefficients independent of $\Delta$;
see~\cite{octant1,Indumathi:2004kd,octant} for details. From the
last line in Eq.~\ref{eqP}, we see that, since {$A$} is positive for $\nu$
and negative for $\overline{\nu}$, hence $P^{m}_{\mu\mu}$ is sensitive
to the sign of the 2--3 mass squared difference via $(A - \Delta)$,
or the neutrino mass ordering. A similar dependence is seen in
the oscillation probability $P_{e\mu}$ as well. Note that the above
expressions were given in order to clarify the dependences on the various
oscillation parameters; precise numerical computations for the oscillation
probabilities are used in the results section.

Before we study upward-going muons for their sensitivity to neutrino
oscillations, we present some details on the muon detector resolution
simulation studies as well as the backgrounds to the process in the next
two sections. There are three different sources of muons in ICAL. These
are

\begin{enumerate}

\item {\bf Cosmic ray muon events}, where cosmic muons enter the detector
from above; these constitute a major background to both the other events,
{\em viz.},

\item {\bf Standard muon events}, where muons arise from CC interactions
of atmospheric $\nu_\mu$ ($\overline{\nu}_\mu$) in the detector, with
the neutrinos entering the detector from all directions.

\item {\bf Rock muon events}, where the muons arise from CC interactions
of atmospheric $\nu_\mu$ ($\overline{\nu}_\mu$) with the rock
surrounding the detector; the associated hadrons produced in the
interaction are absorbed by the rock and only the muons reach the
detector. While these can enter the detector from all directions, they
are indistinguishable from cosmic muons entering in the downward
direction; hence only upward-going rock muons can be distinguished. We
discuss more details about backgrounds in section~\ref{back}.

\end{enumerate}

\section{Detector Response for Muons}
\label{det_res}

\subsection{GEANT4 simulation and track reconstruction}

\begin{figure}[htp]
\centering 
\includegraphics[width=0.55\textwidth]{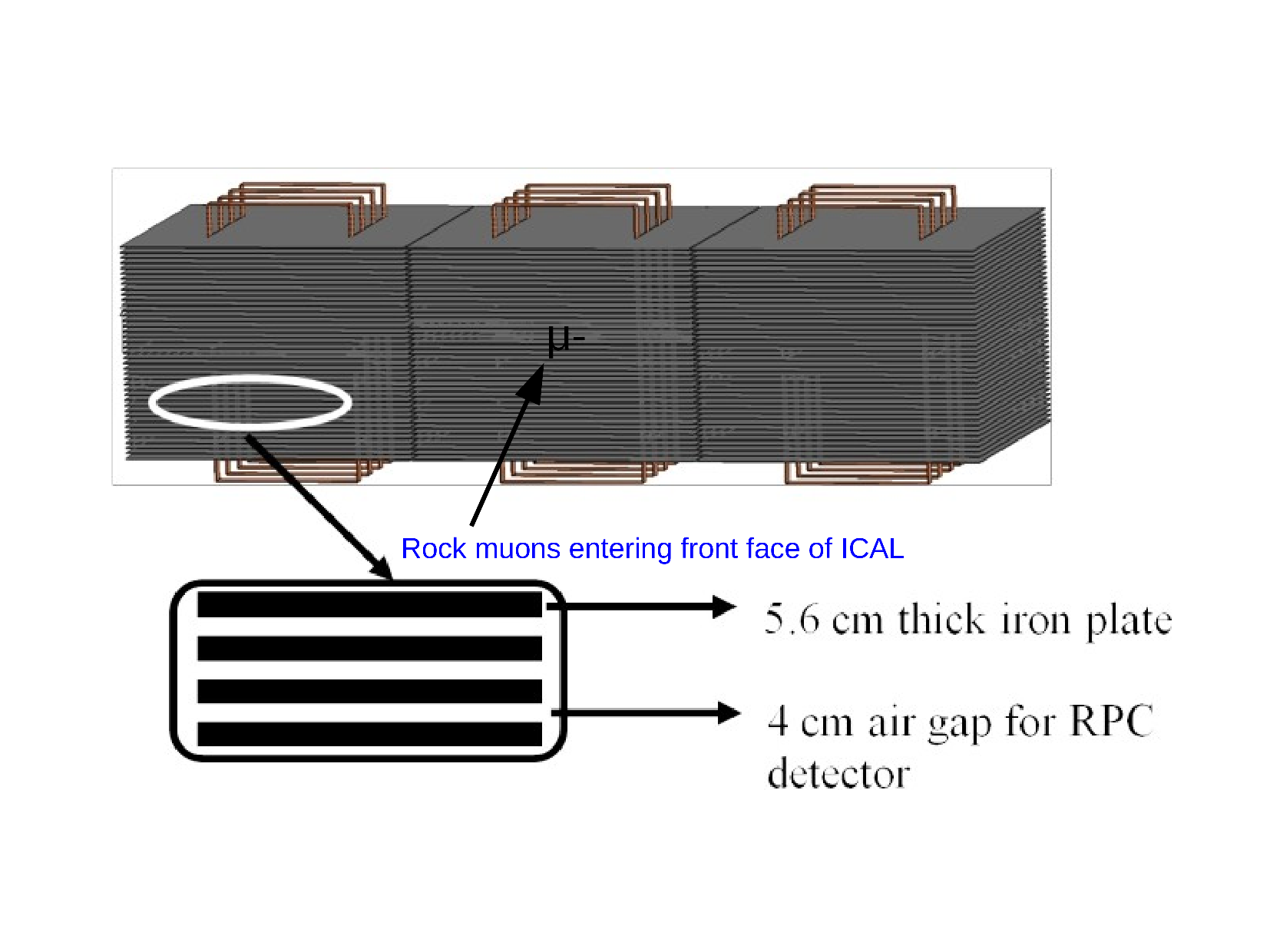}
\caption{Schematic of the proposed ICAL detector. Shown are the 3
identical modules of dimensions $16 \times 16 \times 14.45 \hbox{
m}^3$. The inset shows the 4 cm gaps where the active detector
elements, the RPCs will be inserted. Note that the rock muons entering the front face of the ICAL detector has been shown.}
\label{fig:ICALschematic}
\end{figure}

The proposed magnetized ICAL detector at INO with 1.3--1.5 Tesla
magnetic field will be capable of distinguishing $\mu^{+}$ and $\mu^{-}$
which arise from CC interactions of $\overline{\nu}_\mu$ and $\nu_\mu$
respectively. The ICAL detector as simulated in the ICAL GEANT4 code
\cite{geant} that consists of three identical modules of
dimension 16 m $\times$ 16 m $\times$ 14.45 m. Each module consists of
151 alternate layers of 5.6 cm thick iron plates sandwiched between glass
Resistive Plate Chambers (RPCs)~\cite{satya}, having total mass of about
51 kton; see figure~\ref{fig:ICALschematic}. The total number of RPCs for
ICAL will be 29,000, each having dimensions of $2 \hbox{ m} \times 2 \hbox{
m}$ in the $x$-$y$ (horizontal) plane and 35 mm thick, inserted in
the 45 mm air gap between the plates.

The RPCs consist of two 3 mm thick glass plates of size approximately $2 \times 2$
m$^2$, separated by a 2 mm gap in which
suitable gas flows. Copper pick-up strips of width 28 mm are mounted
on either side of the glass plates, and transverse to each other, so
that there are 64 strips per RPC, above and below the glass plates.
A high voltage of about 10 KV is applied across the glass
plates. When a charged particle passes through an RPC, it creates a
discharge which generates electrons and ions that flow towards the
electrodes. The electrons are
picked up by the pick-up copper strips above and below the RPC and detected
by the associated electronics as a pulse with nano-second timing. The
pick-up strips above and below are in transverse directions thus giving
the $x$ and $y$ locations of the pulse, also called a `hit', while the
RPC layer number in which the pulse was detected gives the $z$ location
of the pulse. In this way, each pulse due to a charged particle is stored
as a hit, with position information in pixels of size in $(x, y, z) =
(2.8, 2.8, 0.2)$ cm, along with the timing information; here $\theta$
= $0^\circ$ corresponds to the vertical direction while $\phi=0$
defines the $x$ axis which is taken to lie along the larger (48 m)
length of the detector. In particular, due to the configurations of
the pick-up strips, the $x$-$z$ and $y$-$z$ information are separately
available. This information is used to determine the momentum of the
muon. This is discussed later.

\begin{figure}[htp]
\centering 
\includegraphics[width=0.49\textwidth]{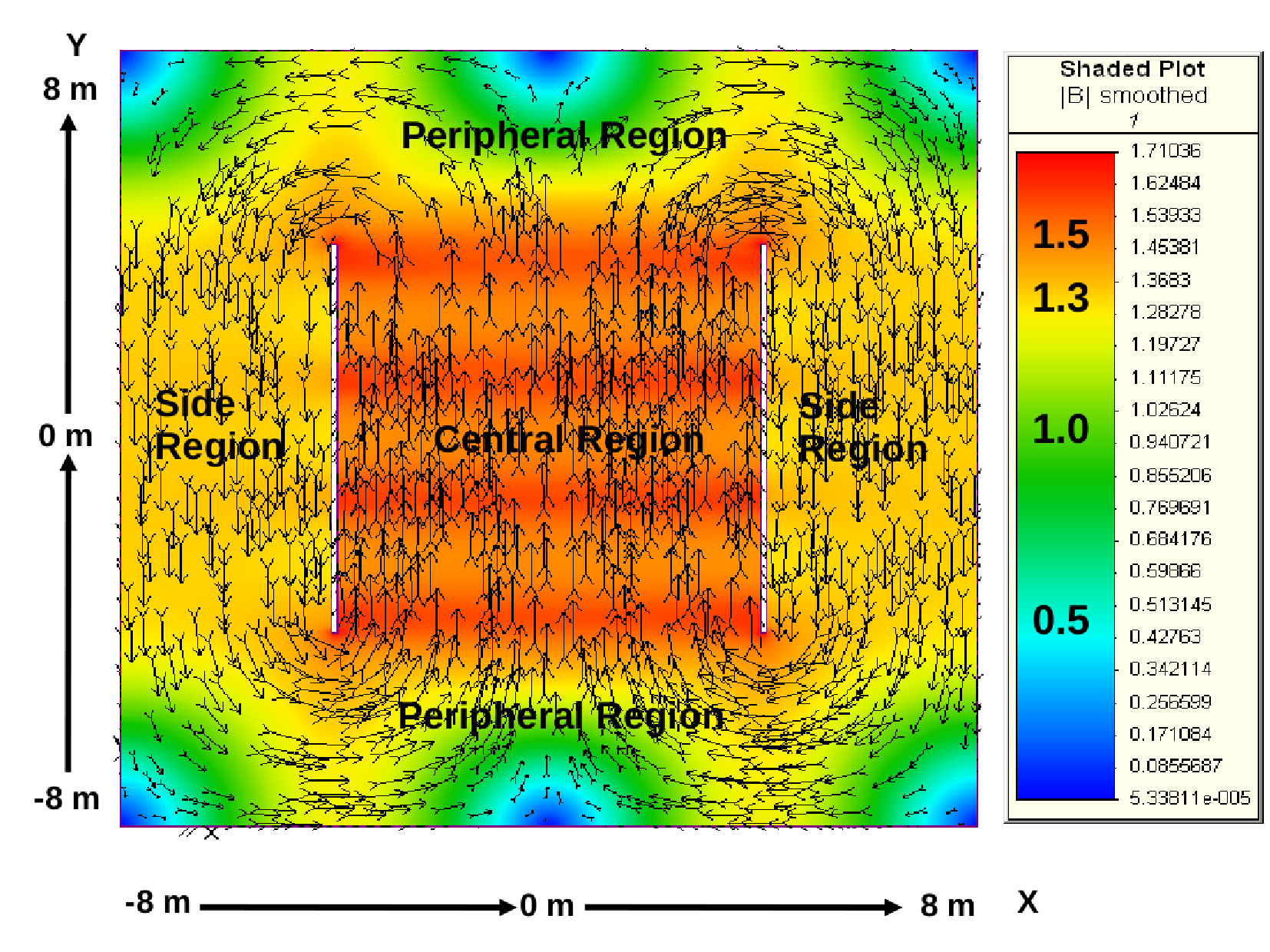}
\caption{Magnetic field map in the $x$-$y$ plane as generated by the MAGNET6
software in the central iron layer of the central
module~\cite{magnetcode}; the gaps correspond to the slots for the
copper coils to pass through.}
\label{fig:magfield}
\end{figure}

The magnetic field is generated by passing current through copper
coils, which pass through coil slots in the plates as shown in figure~\ref{fig:magfield}. It can be seen that due to the coil geometry,
the magnetic field is confined to the $x$-$y$ plane (the plane of the
iron). Furthermore, the magnetic field is distributed in such a way that
it divides the whole ICAL into three regions. The main region is the
``central region'' ~\cite{central} within the coils slots (with $\vert x,
y \vert \le 4$ m in the central module and analogous regions in the outer
modules) which has the highest, as well as most uniform magnetic field
in the $y$ direction, while the magnetic field in the ``side region''
(outside the coil slots in the $x$ direction) is about 15\% smaller
and in the opposite direction. The region labelled as ``peripheral
region''~\cite{peripheral} (outside the central region with $\vert y
\vert \ge 4$ m) has the most varying magnetic field in both magnitude
and direction. Hence the side and peripheral regions will be affected by
having a more complicated magnetic field as well as having edge effects
i.e., both of them have a larger fraction of events (about 23\% compared
to 12\% of events with vertex in the central region) where only a part of
the muon trajectory/track is contained and detected within the detector.

As stated earlier, as the muon passes through the detector, it leaves a
hit in each RPC that it traverses. Due to the presence of the magnetic
field, the path of the muon is bent and hence the succeeding hits form
a curved `track' in the detector. Events are analyzed if there are at
least 3 hits, and the hits are sent to a Kalman filter to determine the
muon charge-sign and momentum. The Kalman filter uses the knowledge of
the local magnetic field to try and fit a
``track'' to multiple sets of hits. Recall that the mutually transverse
pick-up strips yielded information of hits in the $x$-$z$ and $y$-$z$
planes which can be combined to give $x,y,z$ information. The charge
to momentum ($q/p$) ratio for each track is then determined by iterating the information
of the vector containing the location of the hits, the slopes and $q/p$
ratio $(x,
y, z, dx/dz, dy/dz, q/p)$. The initial values of the location are those
of the first hit; the slopes obtained from the first two hits, and
$q/p$ set to zero initially. The GEANT4 ICAL simulation uses the
magnetic field map and the detector geometry to construct a gain matrix
that predicts the location of the next hit in the adjacent layer. Once
it finds such a hit, it adds it to the track and carries on until it
has accumulated a set of hits into a track. There can be more than one
such track; the longest one is identified as the muon track. From the
extent and direction of bending in the magnetic field, the muon momentum $p$ and the
sign of its charge $q$ is determined. Notice that the magnetic field is
mostly (in the central and side regions) along the $\pm y$ direction;
hence a charged particle travelling purely along the $y$ direction
(azimuthal angle $\phi = \pi/2$) will not experience any magnetic
force. A rock muon travelling upwards into ICAL will have $\cos\theta >
0$ and hence the $z$-component of its momentum, $p_z \ne 0$. In general,
the in-plane components of the momentum are also non zero. This ensures
that there is a force that bends the track both in the $x$ (due to $p_z$)
and $z$ (due to $p_x$) directions, thus enabling reconstruction of both
$(\cos\theta, \phi)$ (See Ref.~\cite{peripheral} for details).

\subsection{Fiducial volume and nature of tracks}

Tracks are distinguished based on whether the track is completely or
partially contained inside the detector, as well as whether the track
starts from well within the detector or near the edges. However, note
that standard muon events from CC interactions occurring near the edge
of the detector may be mis-identified as either cosmic muon or rock muon
events. To overcome this ambiguity\footnote{This arises mainly because
ICAL is a detector with layers rather than having a single volume such as in
Super-Kamiokande, where the fiducial volume is actually a part of the
volume of the entire detector.},
the fiducial volume is enumerated as follows: All events
starting from, or produced in the bottom layer and in a region within
50 cm of the four (front, back, left, right) faces of the detector are
considered to be rock muon events. There are rock muon events entering
from above (both in the top layer and from the four sides) but they are
lost in the huge cosmic muon background which is orders of magnitude
larger and are ignored. Hence these rock events will contain a small
fraction of standard muon events.

The following possible type of tracks exist:

\begin{enumerate}

\item The track is completely contained within the (fiducial volume of
the) detector. This indicates that a CC interaction occurred at the
vertex, producing a muon that stopped within the detector. This is
identified as a {\em completely contained atmospheric neutrino or
standard muon event.}

\item The vertex of the track is completely contained within the (fiducial
volume of the) detector, but the track may itself not be fully
contained. This indicates that a CC interaction occurred
at the vertex, producing a muon that exited the detector. This
is identified as a {\em partially contained atmospheric neutrino or
standard muon event.}

\item The track starts from outside the detector (the first ``hit" is
outside the fiducial volume) and stops inside the detector, moving in
the upward direction. This is identified as a {\em partially contained
rock muon event}. A sample track for such an event, entering the detector
from below with $E=5$ GeV, $\theta=30^\circ$ and $\phi=60^\circ$ is
shown in figure~\ref{fig:track5_3D} as a function of time, while figure~\ref{fig:track5} shows the $x$-$z$ and $y$-$z$ projections of the
original and corresponding digitised track.

\begin{figure}[htp]
\centering 
\includegraphics[width=0.5\textwidth]{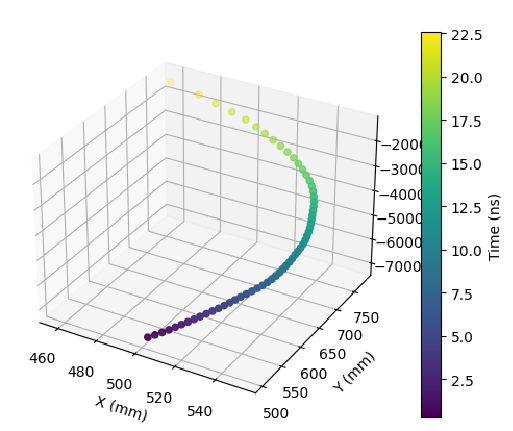}
\caption{Sample track of a 5 GeV rock $\mu^+$ with initial angle $(\theta,
\phi) = (30^\circ, 60^\circ)$ starting at the bottom of the detector in
the central module.}
\label{fig:track5_3D}
\end{figure}

\begin{figure*}[thp]
\centering 
\includegraphics[width=0.49\textwidth]{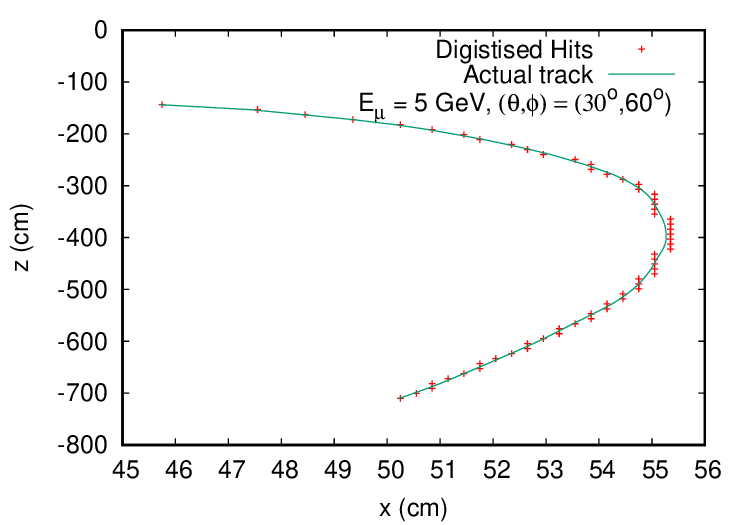}
\includegraphics[width=0.49\textwidth]{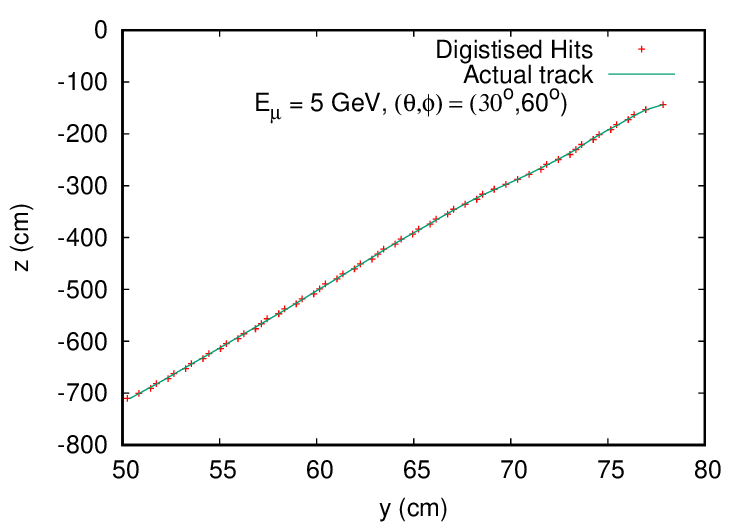}
\caption{The $x$-$z$ (lfet panel) and $y$-$z$ (right panel) projections of the track shown in figure~\ref{fig:track5_3D} and the corresponding digitised ``hits" in $3
\times 3$ cm$^2$ pixels shown in green.}
\label{fig:track5}
\end{figure*}

Notice that the muon starts out moving in the positive $x$ direction,
from the bottom of the detector ($z=-714$ cm) somewhere in the central
region of the central module, where the magnetic field is dominantly
in the positive $y$ direction; see figure~\ref{fig:magfield}. Since the
muon has momentum components in both the positive $x$ and positive $z$
directions, the force on the (positively charged) muon due to the magnetic
field bends it in the positive $z$ and negative $x$ directions. Hence the
muon track, which had started out in the positive $x$ direction, bends
around towards the negative $x$ direction and continues to move upwards;
see the $x$-$z$ projection of the track in figure~\ref{fig:track5}. There
is no force here in the $y$ direction and hence the track in the $y$-$z$
plane is practically a straight line.

\item The track starts from outside the detector (the first ``hit" is
outside the fiducial volume) and exits the detector, moving in
the upward direction. This is identified as a {\em through-going
rock muon event.} Such a sample track is shown for a 200 GeV muon
entering from the bottom in the central region of the central module,
with $(\theta, \phi) = (0^\circ, 0^\circ)$ in figure~\ref{fig:track200_3D}.
Since there is no initial momentum in the $y$ direction, the small
changes in $y$ are due to local scattering in the detector. However, on
digitisation, the track is seen to be constant in the $y$ direction,
bending to the left in $x$, as expected, and exiting the detector from
the top, as can be seen in figure~\ref{fig:track200}.

\begin{figure}[thp]
\centering 
\includegraphics[width=0.5\textwidth]{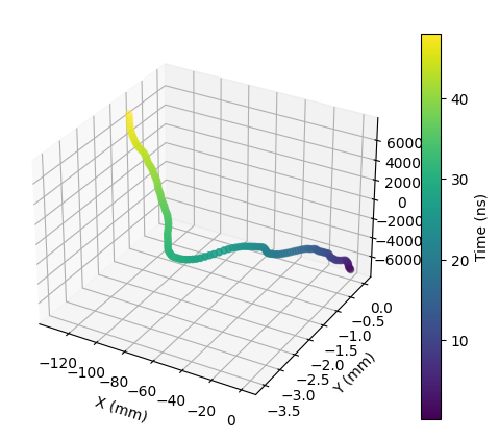}
\caption{As in figure~\ref{fig:track5_3D} for a 200 GeV muon, with
$(\theta, \phi) = (0^\circ, 0^\circ)$ starting at the bottom of the
detector in the central module. Notice the very small scale in $y$.}
\label{fig:track200_3D}
\end{figure}

\begin{figure*}[thp]
\centering 
\includegraphics[width=0.49\textwidth]{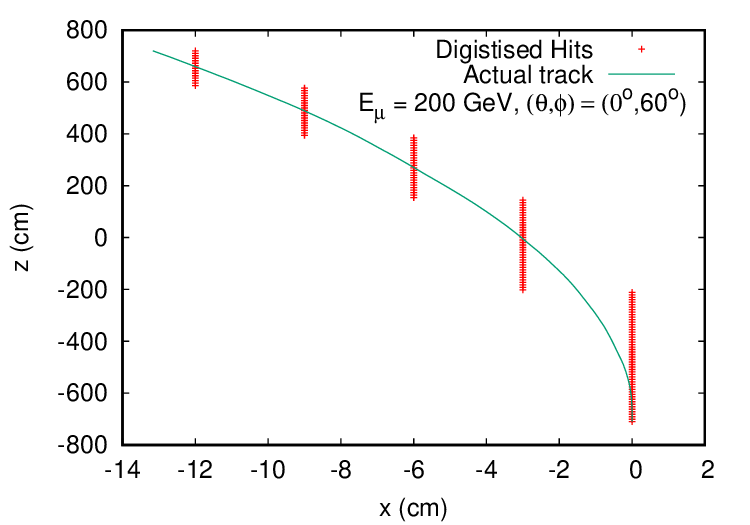}
\includegraphics[width=0.49\textwidth]{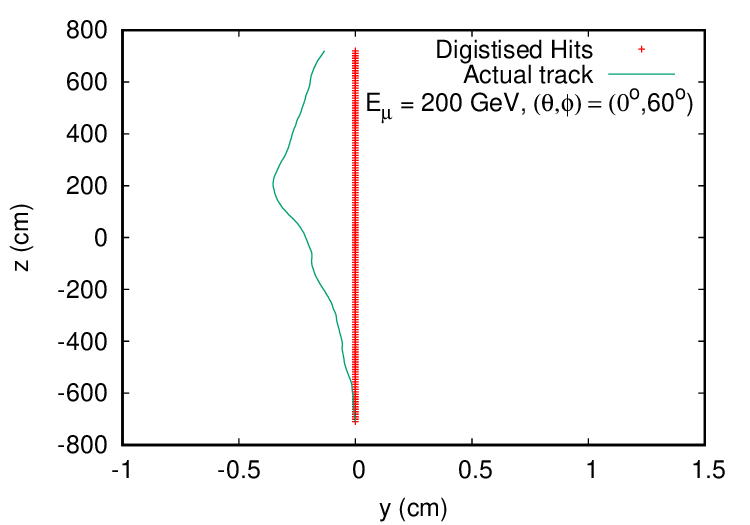}
\caption{The $x$-$z$ (left panel) and $y$-$z$ (right panel) projections of the track shown in figure~\ref{fig:track200_3D} and the corresponding digitised ``hits" in $3 \times 3$ cm$^2$ pixels shown in green.}
\label{fig:track200}
\end{figure*}

\item When the track starts from outside the detector and is moving in
the downward direction, this is identified as a {\em cosmic muon event.}

\end{enumerate}

\subsection{Muon track reconstruction}

As has been discussed in~\cite{central,peripheral}, ICAL has good energy
and direction resolution in both the central and peripheral regions
of the detector.
However, these studies on muon energy, direction
reconstruction and charge identification capability were performed with
a view to understand the detector response for the main events at ICAL,
viz., CC muon neutrino interactions inside ICAL. Hence, all these studies
used a simulated data sample where the neutrino interactions occurred {\em
inside} ICAL so that the produced muons were also often inside the detector.

For the current study, we need to understand the response of ICAL to
muons that are entering the detector {\em from outside}. These are also
very ``clean'' events in that there are no accompanying hadrons \cite{hadronresponse}. Hence
we first calibrated the detector response to such events using a
GEANT4-based \cite{geant} code to generate and propagate the events
through the simulated ICAL detector. Note that the upward-going muons
enter the detector through five different faces (left, right, front,
back and bottom). While about half the upward-going muons have their
vertices in the bottom of the detector, the four sides (left, right,
front and back) of ICAL together account for the other half of the events.
In each case, the muon experiences a different local magnetic field. Hence
the response will be different in each case. However, muons that enter
through the bottom face of the detector experience regions corresponding
to all the possible choices---central, peripheral and side. Hence, we
study the response of the bottom face of the whole ICAL to muons. This
contains portions of the ``central'' and ``peripheral'' regions and so
its response is likely to be intermediate between the two.

As the muon enters ICAL, it produces signals in the RPCs. These
signals are localized to a size of 3 cm, which determines the spatial
resolution of the muon track in the $x$- and $y$-directions and are
called ``hits''. An event has hits in several layers (along the
$z$-direction). Since the detection efficiency of the RPCs is 95\%,
there may be different numbers of $x$- and $y$-hits in any layer. Hence
the total hits $N_{hits}$ per event are then determined
as the sum of the maximum value of the $x$- or $y$-hits for each layer.
The entire set of hits in the event is passed to a Kalman filter
algorithm that selects out the hits associated with the muon track,
while simultaneously fitting the track to reconstruct the momentum,
direction, and sign of charge, based on propagation in the local
magnetic field. The hits that are rejected by the algorithm constitute
the so-called ``hadron hits'' that are used to calibrate the associated
hadron energy. Since at least three hits in two layers were required to
identify a hadron shower, it was found \cite{ICAL:2015stm} that only
hadrons with energy $E_\mu > 1$ GeV formed showers.

The selection criteria to choose or drop an event were decided so
as to get reasonable fits and hence resolutions. Mainly one major
selection criterion has been applied in the region to remove low
energy tails; this was similar to that used in studying the peripheral
muons~\cite{peripheral}. The events were selected in a manner such that,
if the track was partially contained and ended well within the detector
(most likely scenario at lower energies) then it was considered for
analysis, but if it was through-going (more likely at higher energies
greater than about 10--15 GeV), then the event was selected only if
$N_{hits} > 15$. A somewhat looser constraint has been obtained by also
taking into account the angle at which the muon enters (since the number
of layers traversed and hence the number of hits in a track are dependent
on this) by demanding that $N_{hits}/\cos\theta > 15$. This criterion
removed muons that exited the detector leaving very short tracks inside,
which were typically reconstructed with much smaller momenta than the
true values.

\begin{figure*}[thp]
  \centering
\includegraphics[width=0.49\textwidth]{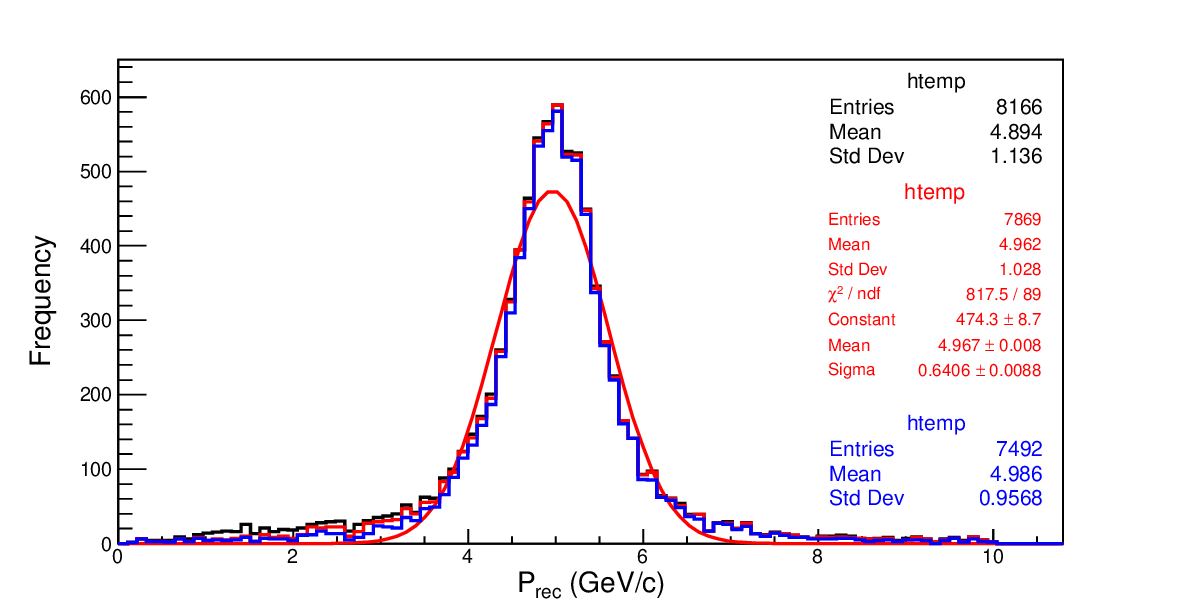}
\includegraphics[width=0.49\textwidth]{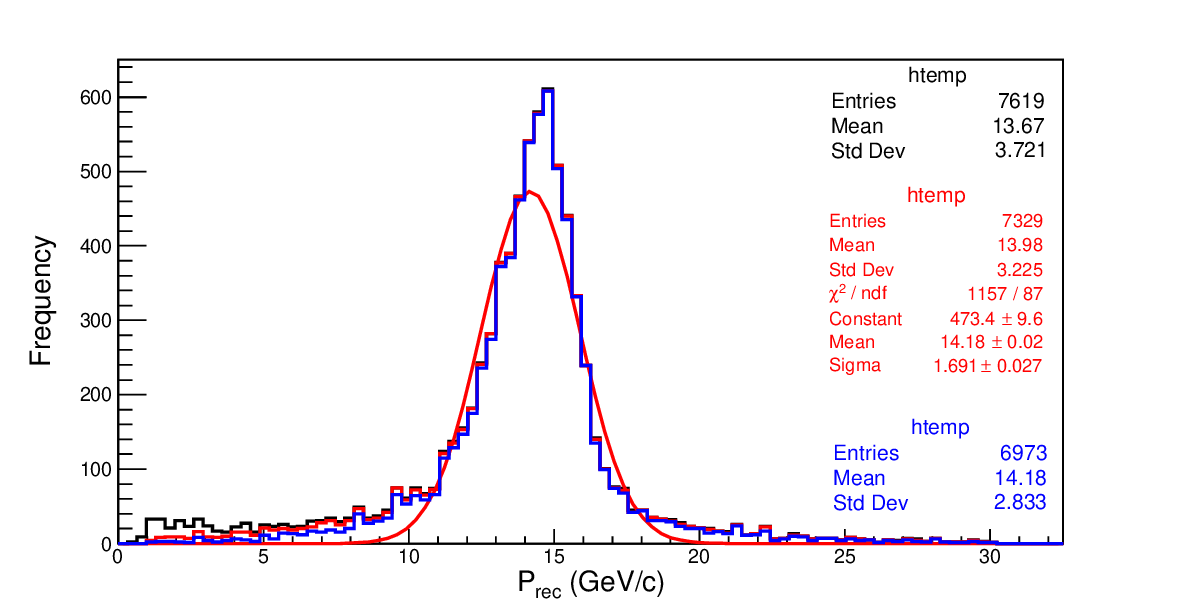}
\caption{The reconstructed momenta $P_{rec}$ using selection criteria
$N_{hits}>n_0$ for through-going events in the
bottom region of ICAL at: ($P_{in}$, $cos\theta$) = (5 GeV/c, 0.65) (left
panel) and ($P_{in}$, $cos\theta$) = (15 GeV/c, 0.65) (right panel). In
both figures, the black curve is without constraints on $N_{hits}$,
red is with $N_{hits}/\cos\theta > n_0$ and blue is for $N_{hits} >
n_0$; $n_0=15$.  The Gaussian fit to the last histogram is also shown.}
\label{fig:cuts}
\end{figure*}

This can be seen from figure~\ref{fig:cuts} which explains the effect
of the selection criterion on reconstructed momentum ($P_{rec}$) in
the region. Here, 10,000 muons with fixed input momenta and direction
($\cos\theta$) were randomly generated with vertices uniformly distributed
on the bottom face of ICAL, with uniform random azimuthal angle, $0 <
\phi < 2\pi$ (smeared).

It can be seen that, at lower energy ($P_{in}$ = 5 GeV/c), the $N_{hits}$
criterion does not significantly affect the momentum distribution as most
of the events are fully contained. On the other hand, the hump at lower
energy for $P_{in}$ = 15 GeV/c is due to the charge mis-identification
which has been eliminated with the $N_{hits}$ cut (In fact, the condition
that only one track be reconstructed, as demanded in the peripheral muon
analysis~\cite{peripheral}, was not required as the present constraint
on $N_{hits}$ was found to be sufficient). The resulting histogram was
fitted with a Gaussian distribution to determine its width $\sigma$,
from which the muon momentum resolution has been defined as the ratio
of the width to the initial momentum ($\sigma/P_{in}$).

Note that 3 layers is the criterion for the events to be passed to
the Kalman filter. All rock muon events have tracks starting from
the edges or faces of the detector and hence are typed as partially
contained events (defined as at least one end of the track being close
to any edge of the detector). For all such events, the criterion is
$N_{hits}/\cos\theta \ge 15$. For vertical events, with $\cos\theta =
1$, this leads to a sufficiently long path length traversing about 15
layers. Since the separation between the RPCs is nearly 10 cm, this
corresponds to a time difference between the first and last hits of
$\Delta T = 5$ ns. Even for horizontal angles such as $\cos\theta = 0.2$
($\theta \sim 80^\circ$), where the muons cross only 3 layers, such a
slant track would traverse a longer length of $9.6/\cos\theta \sim 0.5 m$
{\it through each layer}. So even if the total number of layers is less,
this criterion would correspond to a total distance of about 1.5 m or
$\Delta T = 5$ ns, which is $5\sigma$ of the 1 ns timing resolution of
the RPCs. In short, the criterion used ensures that the time difference
$\Delta T$ between the first and last hits is sufficiently large for
unambiguous up/down discrimination.  We will later see the importance
of this selection criteria in reducing the cosmic muon background.

\begin{figure}[htp]
\centering
\includegraphics[width=0.5\textwidth, height=0.3\textwidth]{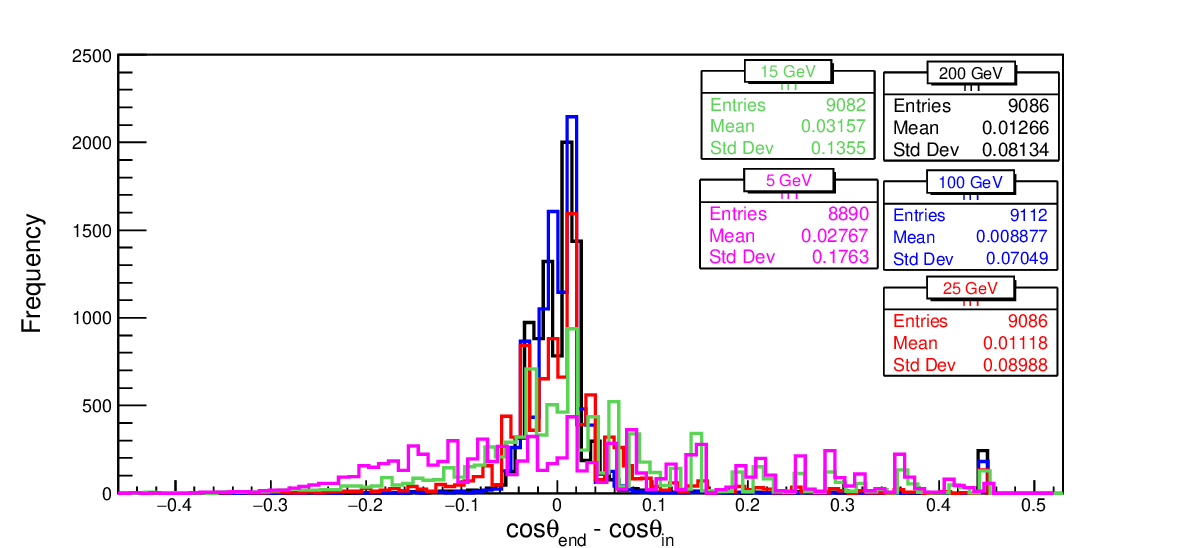}
\caption{Sample distributions of $\cos\theta_{end} -
\cos\theta_{in}$ for different initial values of muon energy, for
$\cos\theta = 0.55$ and $-\pi \le \phi \le \pi$.}
\label{fig:bending}
\end{figure}

The bending of the tracks determines not just the magnitude of the
momentum of the muons but their direction as well.
Figure~\ref{fig:bending} shows the histograms of the difference of
the zenith angle at the start and end of the track for sample muon
energies, for an input angle $\cos\theta = 0.55$ and $-\pi \le
\phi \le \pi$.

It is seen from the mean and rms values that the distribution is more
or less symmetric, especially at higher energies. This is because the
magnetic field changes direction outside the coil slots and hence the
direction of bending switches when the particle goes from a region inside
the slots to a region outside (or vice versa). This is more likely to
occur at higher energies when the muon traverses the entire detector
and even exits it. Hence the smearing of events over the azimuthal angle
dilutes the effect.

\begin{figure*}[htp]
\centering
\includegraphics[width=0.49\textwidth]{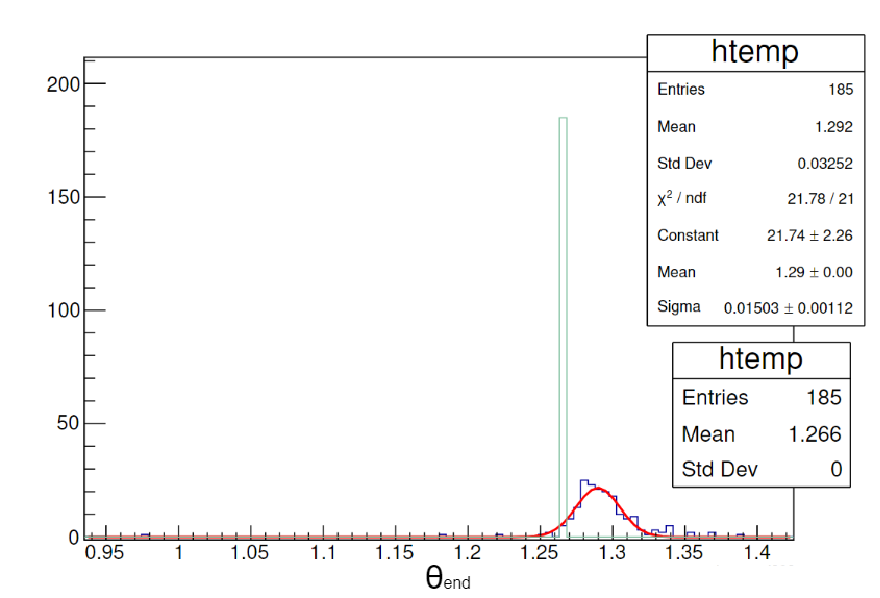}
\includegraphics[width=0.49\textwidth]{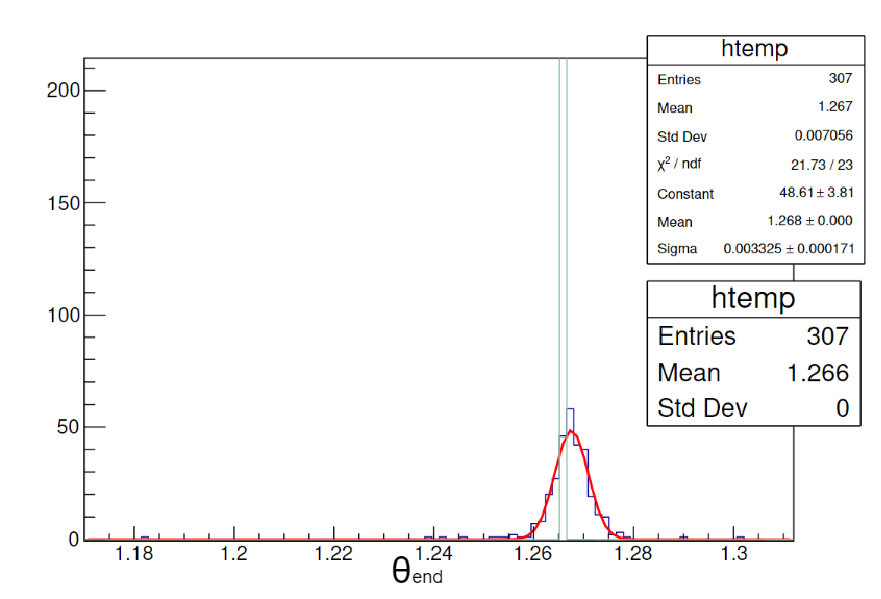}
\caption{$\theta_{end}$ distributions for $\cos\theta_{in} = 0.55, \phi
\le 1$, for $E_\mu=10$ GeV (left), 100 GeV (right).  The green (blue)
histogram corresponds to the input (end) zenith angle.}
\label{fig:fixbend}
\end{figure*}

For better clarity, the zenith angle at the start and end of the track
are plotted in figure~\ref{fig:fixbend}.
Shown are the $\theta_{end}$ distributions for two sample energies, a lower
energy of 10 GeV and a higher one of 100 GeV for fixed value of input
zenith angle, $\cos\theta = 0.55$, generated in the central region of the
detector and so the magnetic field is initially in the $+y$ direction.
Here the events with azimuthal angle $\phi \le 1$ are selected so that
the component of the particle momentum along the $x$ axis is always
positive. It can be seen that the reconstructed zenith angle is always
larger than the input value because negative muons were generated.
This separation is much less at 100 GeV than at 10 GeV. This
bending allows for the reconstruction of the muon momentum and the
sign of its charge. As can be seen from the sample tracks shown in figures~\ref{fig:track5_3D}, \ref{fig:track5}, \ref{fig:track200_3D}
and \ref{fig:track200}, the average values of $(\cos\theta_{in} -
\cos\theta_{end})$ are not really reflective of the detailed bending of
the track which is complex and depends on the charge-sign of the muon,
the (variable) magnetic field components in the local region of the track,
and the momentum components of the incoming muon. All these affect the
reconstruction and charge identification efficiency and the muon
momentum resolution, which we discuss in the next section.

\subsection{Reconstruction Efficiency and Resolution of Muons}
\label{res_eff}

\begin{figure*}[bhp]
\centering 
\includegraphics[width=0.49\textwidth]{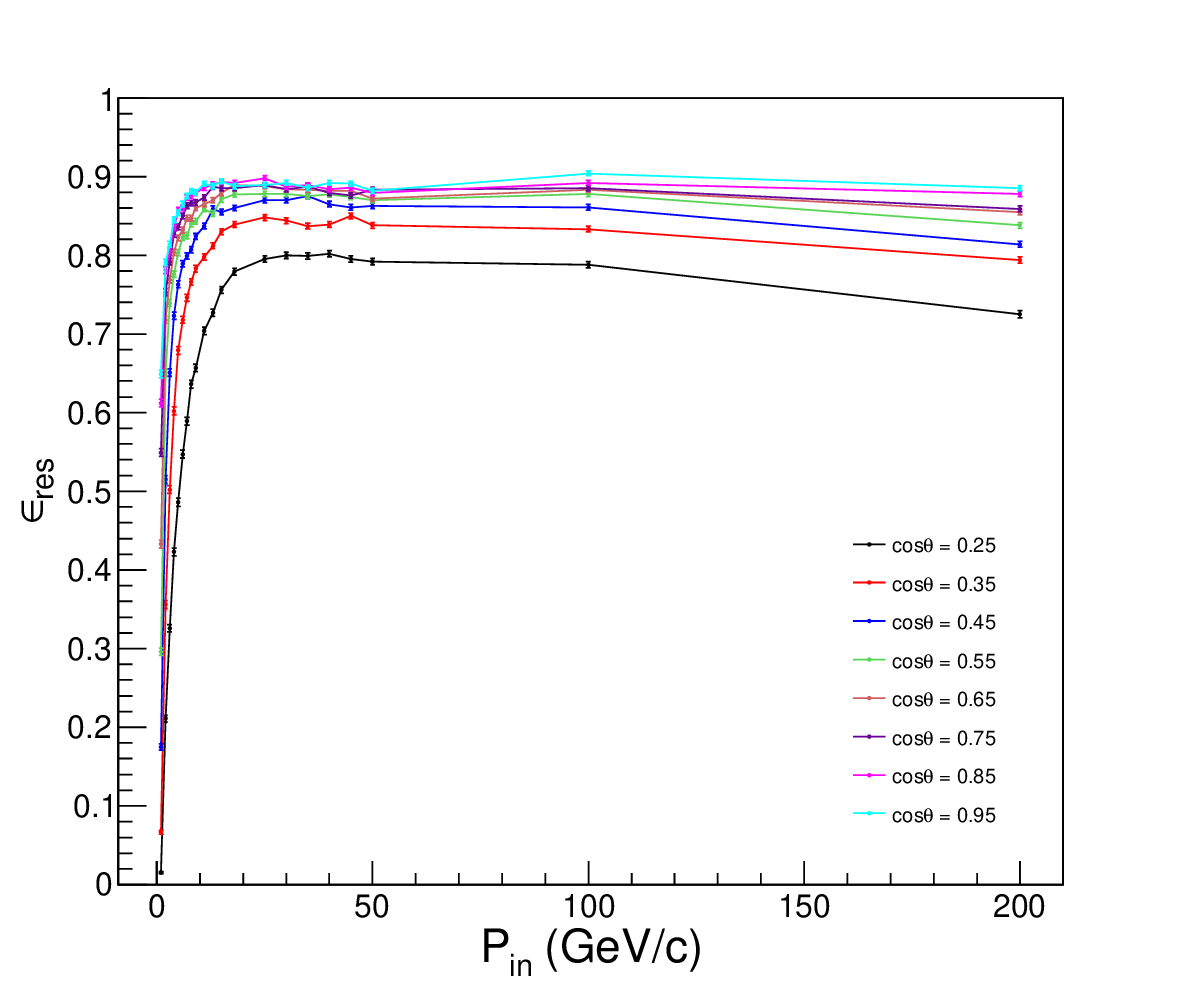}
\includegraphics[width=0.49\textwidth]{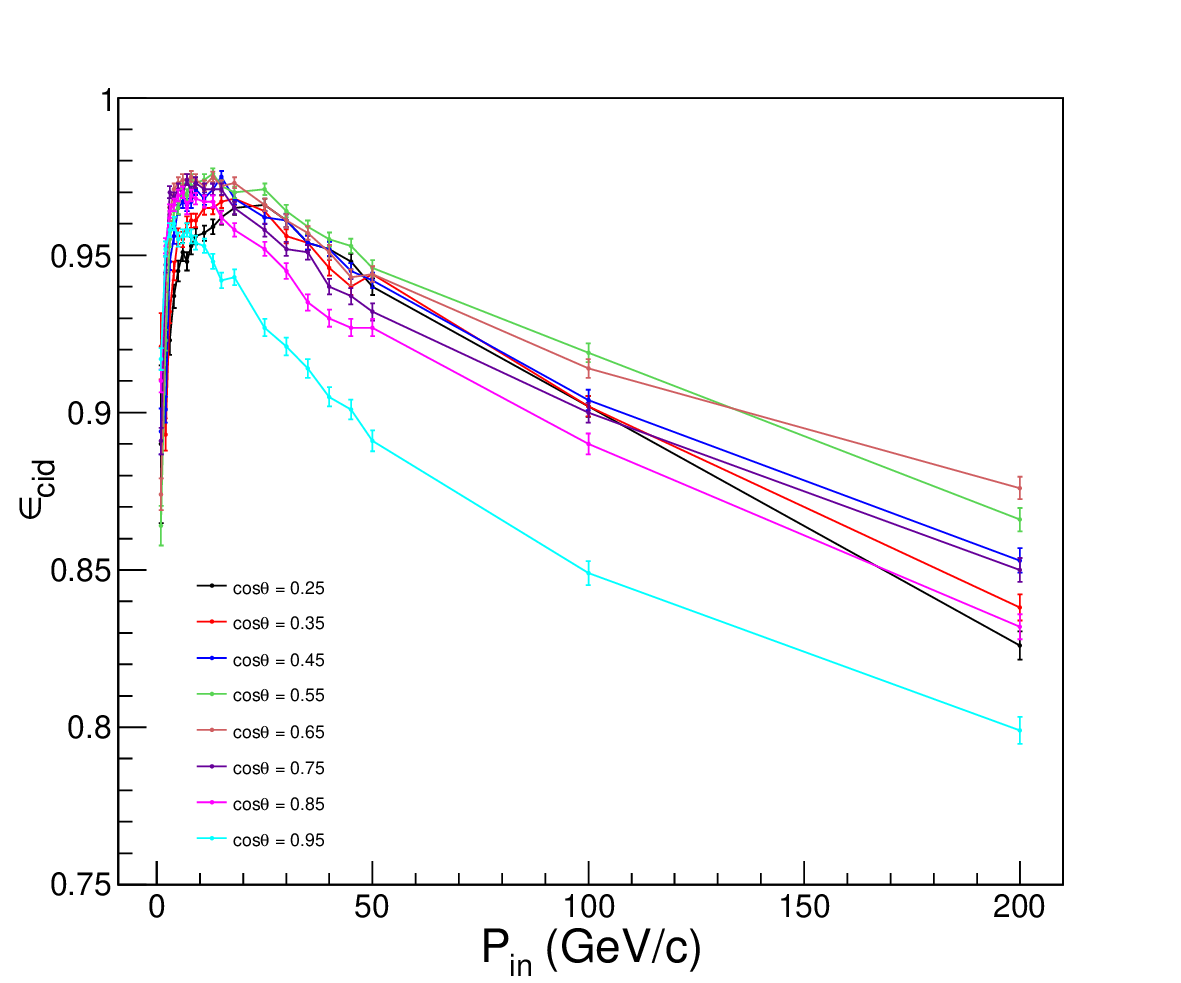}
\caption{The reconstruction efficiency averaged over azimuthal angle
$\phi$ for $N_{hits}/\cos\theta > 15$, as a function of $P_{in}$ for
different zenith angles, $\cos\theta = 0.25, \cdots, 0.95$ (left panel). The charge
identification efficiency averaged over $\phi$ for
different zenith angles, $\cos\theta = 0.25, \cdots, 0.95$ (right panel). (Note that the
$y$-axes scales are different.)}
\label{fig:eff}
\end{figure*}

Now we discuss the results on reconstruction efficiencies and energy
and angular resolutions for the muons based on the selection criteria
discussed earlier. Figure~\ref{fig:eff} shows the reconstruction and charge
identification efficiencies in the bottom region of ICAL.

The reconstruction efficiency, which is defined as the ratio of the
number of GEANT4-simulated events reconstructed to the total events
simulated, was found to be greater than 85\% for energies less than 50
GeV and for angles greater than $\cos\theta > 0.35$. The relative charge
identification (cid) efficiency, which is the ratio of the number of
events with correctly identified muon charge sign to the total number
of reconstructed events, is better than 95\%, and in fact nearly 97\%
for $p_\mu < 20$ GeV; it is better than 85\% for $E_\mu$ upto 150 GeV
and $\cos\theta > 0.35$.

\begin{figure*}[htp]
  \centering
\includegraphics[width=0.49\textwidth]{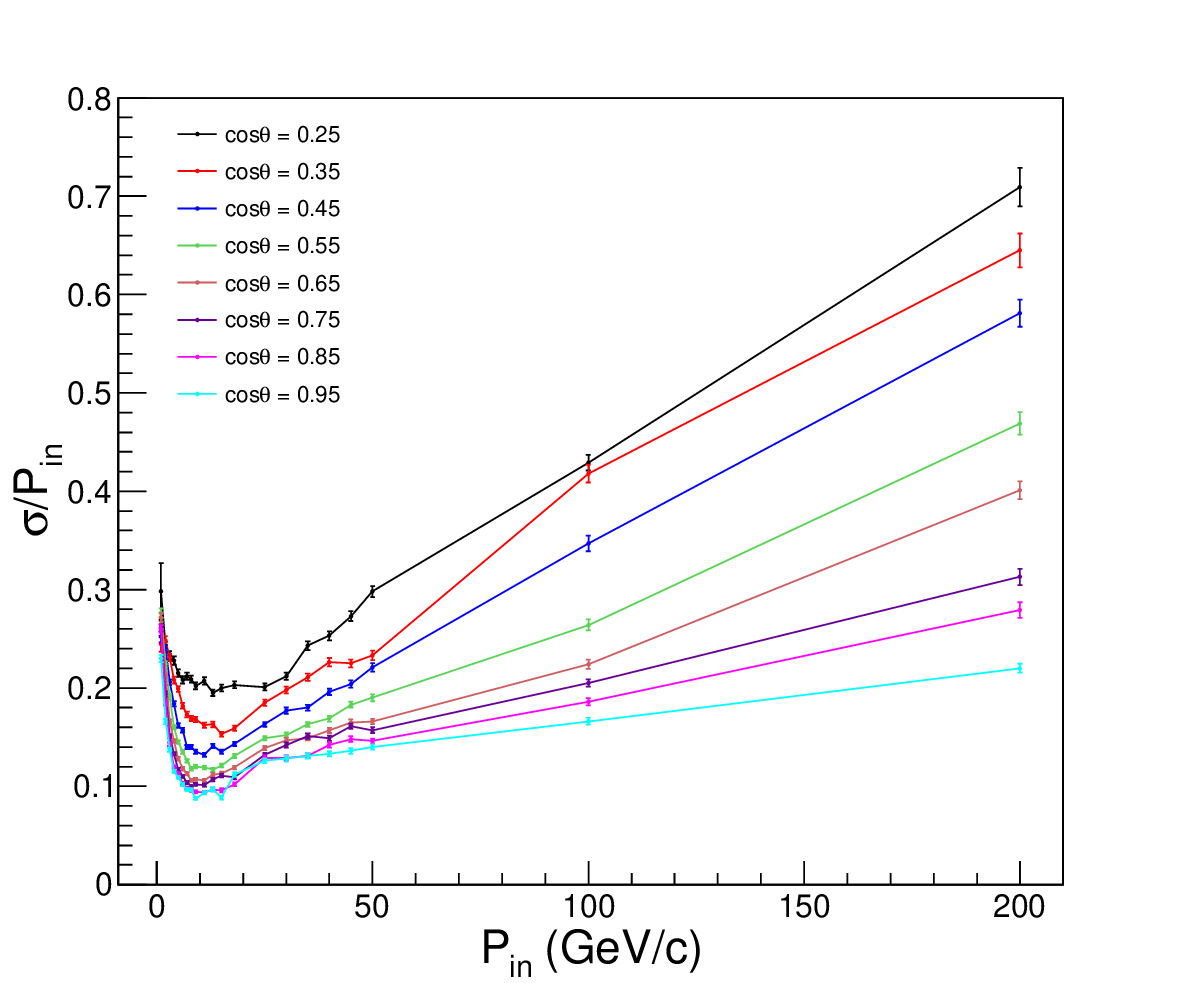}
\includegraphics[width=0.49\textwidth]{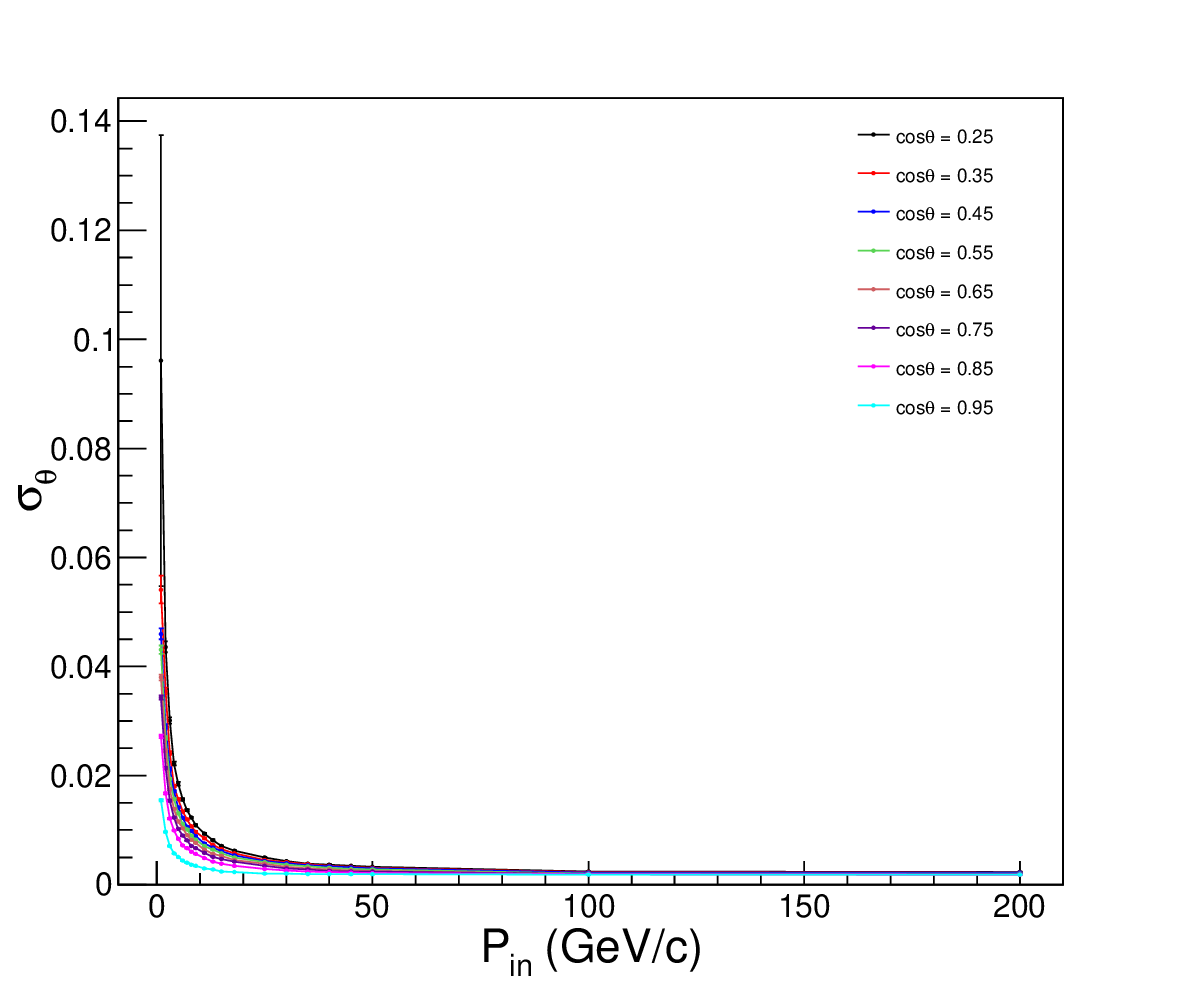}
\caption{Muon momentum resolution as a function of input momentum
$P_{in}$ for different $\cos\theta$, averaged over $\phi$, for
$N_{hits}/\cos\theta > 15$ (left panel). The angular $\theta$ resolution (in radians) is
shown as a function of the input momentum $P_{in}$ for different $\cos\theta$, averaged over $\phi$, for
$N_{hits}/\cos\theta > 15$ (right panel). (Note that the $y$-axes scales are different.)}
\label{fig:resol}
\end{figure*}

Figure~\ref{fig:resol} shows the muon momentum resolution
$\sigma/P_{in}$ and the zenith angle $\theta$ resolution. The detector can optimally detect muons with about
10--12 GeV energy at all angles. Muons with higher energy can exit
the detector and hence the resolution decreases beyond this point. Note that the zenith angle
resolution is the value of the width $\sigma$ of the fitted Gaussian
distributions in radians.

The muon resolutions are better than about 20\% for muon energies
less than 50 GeV and for angles greater than $\cos\theta > 0.35$,
and worsen for larger energies and angles but remain less than
50\% upto $E_\mu < 150$ GeV. The $\theta$ resolution, which was
about a degree for few GeV region, is similar to that obtained from
earlier studies~\cite{peripheral}.  We shall use these values of muon
reconstruction efficiency and resolution in our simulation studies of
upward-going muons in the next sections. Before doing this, we list the
main backgrounds to the process of interest, and methods of reducing
them.

\section{Main backgrounds}
\label{back}

Upward-going rock muons can enter the detector from the bottom of ICAL,
from the front and back, as well as from the sides (left and right
faces). While the events from the side are the least, there are roughly
equal number of events from the bottom and the front-back due to the
detector dimensions of $48 \times 16 \times 14.45 \hbox{ m}^3$.

The primary criterion for identification of these rock muons is
the ability to distinguish up- and down-going muons since the
down-going muons are primarily cosmic ray muons. The RPC
timing is crucial for this purpose; the RPCs that have been
designed and tested by the collaboration have a timing resolution
\cite{ICAL:2015stm} of about 1 ns. Since the vertical spacing between
RPCs is 9.6cm, a minimum of hits in at least three layers is required
to unambiguously determine from the timing information whether the muon
was an up-coming or down-going one; in fact, the Kalman filter algorithm
that identifies and fits the muon tracks requires hits across at least
5 contiguous layers (although one or more layers may not have a hit in
them). Earlier GEANT4-based \cite{geant} simulation studies \cite{central}
by the collaboration have shown that the fraction of tracks which are
reconstructed in the wrong direction (upward tracks being identified as
down-going and vice versa) varies from 1.5--4\% for muons with $E_\mu =
1$ GeV and with zenith angles from $\cos\theta = 0.9$--0.2, with worse
reconstruction for larger angles as expected. This fraction decreases to
less than 0.3\% for large angles of $\cos\theta = 0.2$ when the energy
increases to 2 GeV. Hence the probability of muons with energies $E_\mu >
1$ GeV being reconstructed in the wrong direction is very small and will
be ignored. The additional selection criterion described above plays
an important role.

These upward-going muons are to be discriminated from two other types of
events which form the background to this measurement. First, are those
atmospheric neutrino events that produce muons through CC interactions
inside the ICAL detector and are a part of the main studies of ICAL;
we label them as ``standard muons''.  The other background is due to
the cosmic ray muons. We first consider the ``standard muons''
produced in CC interactions inside ICAL.

\subsection{Standard Muon background}

Rock muon events give rise to muon tracks that start at the edges of
ICAL. Note, however, that (up-going) atmospheric neutrinos producing
muons via CC interactions at the edges of the detector can be mistaken for
rock-muon events since their tracks also begin at the detector edges. The
number of rock muon events depends on the aperture, i.e., the {\em area}
of the surface exposed to these muons. In contrast, the number of CC
``standard muon'' events depends on the {\em volume} of detector in
which they are produced. This background can be significantly reduced
by suitable selection criteria as we describe below.

\begin{figure*}[htp]
\centering 
\includegraphics[width=0.32\textwidth]{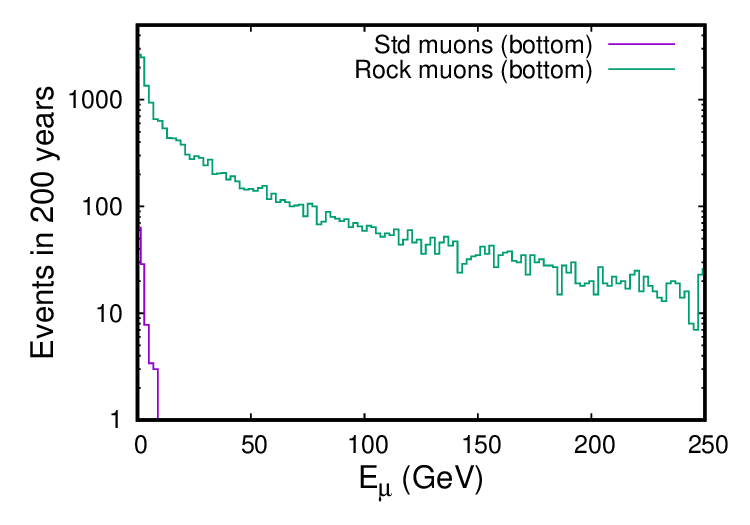}
\includegraphics[width=0.32\textwidth]{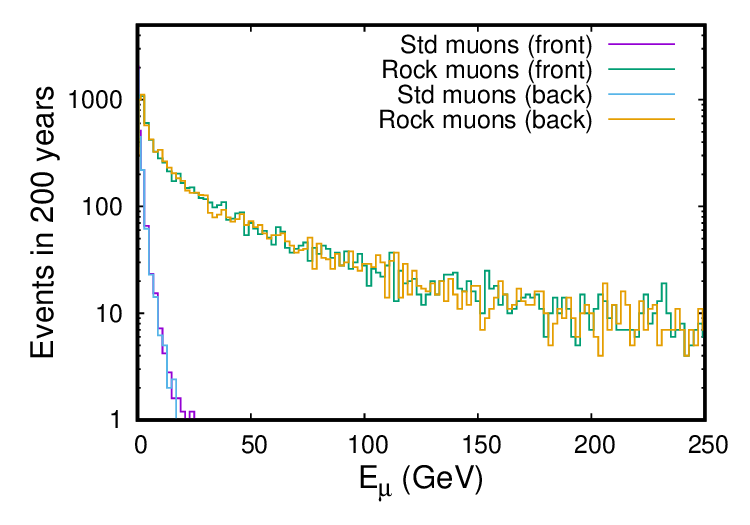}
\includegraphics[width=0.32\textwidth]{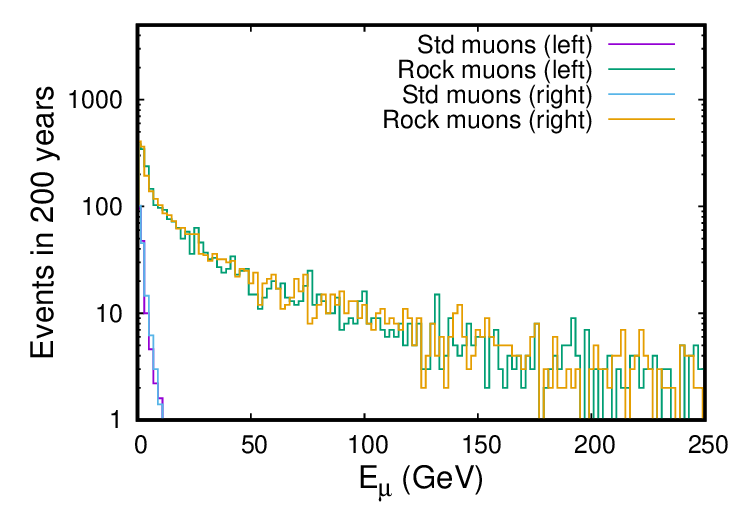}
\caption{Rock events and relevant backgrounds arising from CC
interactions of atmospheric neutrinos for 200 years in ICAL. Shown are
the events that enter from the bottom, front-back, and right-left faces
of the detector respectively. See text for details.}
\label{fig:background}
\end{figure*}

\paragraph{Bottom events}: Figure~\ref{fig:background} shows the
rock events entering from below, through the bottom of ICAL (so-called
``bottom rock events'') that are detected, starting from the bottom-most
RPC layers. Atmospheric neutrinos that {\em interact} with the nucleons
via CC interactions in the bottom-most layer of ICAL also produce muons
that are detected, starting from the bottom-most RPC layer. Among these
events, those which are detected as single tracks (identified as muons)
having no visible accompanying hadronic activity and produced in the
upward direction, can mimic the rock events. Hence such CC atmospheric
muon neutrino events form an irreducible background to the ``bottom rock
events''. These are also shown in figure~\ref{fig:background}. The standard
muon background satisfying the above selection criteria is about 2.5\%
of the bottom-rock events at muon energies of $E_\mu \sim 1$ GeV, and
falls off to 0.5\% ($< 0.2$\%) at $4$ ($> 10$) GeV. We therefore apply a
selection criterion of $E_\mu > 1$ GeV in order to reduce this background
to less than 2.5\%.

\paragraph{Front/Back events}: 
Similarly, the up-going atmospheric neutrinos that produce up-going
muons in CC interactions at the edge of the front and back faces of ICAL
with no visible associated hadronic shower, can mimic rock muon events
arriving through the front and back faces of the detector. Depending on
the direction ($\cos\theta$) of the muon, it can traverse a large distance
before reaching an RPC layer and giving a ``hit'' there. For instance,
a muon at angle $\cos\theta = 0.2$ can traverse $9.6/\cos\theta \sim 50$
cm before reaching an RPC layer. Hence, we have assumed that any muon
track whose first hit is within 50 cm of the detector faces (front, back)
can be considered to be a rock muon event.  This means that all standard
muons produced due to CC neutrino interactions in the iron material
within 0.5 m of the edges form an irreducible background to the rock
muon events. This is in contrast to muons entering from above or below
where they meet an RPC layer just adjacent to the top or bottom iron
layer and hence this irreducible background is relatively larger at low
energies, being 47\%, 11\%, and 2\% of the front- and back-entering rock
muon events at 1, 4, and 10 GeV respectively. The larger background is
because these interactions occur in 150 layers of $48 \times 0.5 \times
0.056 \hbox{ m}^3$ of iron, which is much larger than the interactions
occurring in one bottom layer of dimension $48 \times 16 \times
0.056 \hbox{ m}^3$; see the schematic of the ICAL detector in figure~\ref{fig:ICALschematic}. The rock muons entering the front face 
of the ICAL detector has also been shown. Here, we apply a selection criterion
of $E_\mu > 4$ GeV to reduce this background from the front and back
faces to less than 10\%.

\paragraph{Left/Right events}: 
Lastly, the atmospheric neutrinos entering from the side (left and
right faces) of ICAL can interact via CC interactions to again produce
muons within 50 cm from the detector edges, thus mimicking rock muon
events entering from the sides. These backgrounds amount to 20\%, 3\%,
and $< 1$\% for muons with energies $1, 4$ and 10 GeV. Again we employ
a selection criterion of $E_\mu > 4$ GeV to reduce this background
from the left and right faces to less than 3\%.

\paragraph{Fiducial volume}:
We note that the choice of fiducial volume for rock muon
events of $E_\mu > 4$ GeV muons produced within 0.5 m from the front,
back, left and right edges and the events with $E_\mu > 1$ GeV
produced in the
entire bottom layer, is quite conservative, especially since it is
blind to the direction of the muon, and a smaller
value may cut down the CC neutrino events further. In the absence
of any available studies in this regard, we retain this conservative
choice. In our further analysis, we have chosen to ignore these
background events as being small. It is possible to actually carry out the
detailed analysis including these background events, but that is
beyond the scope of this work.

\subsection{Cosmic Muon Background}

The second kind of background arises from the cosmic ray muon events
produced in the Earth's atmosphere that directly leave tracks in
the ICAL detector. These form the main background to all muon events
(CC atmospheric muon neutrinos producing muons as well as rock muon
events) in the ICAL detector. In particular, those cosmic muons that are
mis-identified in their direction (down-going cosmic muons mis-identified
as being up-coming ones) form the cosmic muon background to rock muon
events. These can be partially eliminated from the rock muon sample
by imposing an angle cut that allows only upward-going muons for the
analysis since there are no cosmic ray muons arriving from below. As a
matter of abundant precaution, we also reject up-going muons that enter
at angles larger than $81^\circ \hbox{ or } \cos\theta < 0.156$.

Note that the decision of an event being up-going or down-going
depends on the ability of the detector to discriminate up and down
events (that is, the probability to reconstruct a true angle $\theta$
as $(\pi - \theta)$ which results in reconstructing $\cos\theta$ with
the wrong sign. In our simulations studies, we have used a sample of 10,000
muons of fixed energy and direction. We have already mentioned earlier
that the selection criteria result in a track length of the muons
which corresponds to a time of 5 ns or more. Since this is 5 standard
deviations away from the RPC resolution of 1 ns, only an event in a
million is expected to be reconstructed in the wrong direction; these
are unobservable in a sample size of 10,000 that we have used for the
analysis. Generating millions of events and analyzing them in GEANT is
a slow and difficult process, and is also limited by the memory space
available on the computer. However, the cosmic muon flux is huge and
so even a small number of these events, mis-identified in direction,
can form a significant background to the rock muon events. In order to
accurately estimate the background from these events, it is therefore
necessary to find the fraction of up-down mis-identified cosmic muon
events using a different approach.

To this end, a Monte Carlo program was written, that generated cosmic
muon events according to the fluxes given in Ref.~\cite{Workman:2022ynf}.
Since the peak height below which INO is proposed to be located is 1.3 km
(approx), and the access tunnel to reach this cavern is 2.1 km long,
the simulation assumed a simplistic conical mountain\footnote{This
reflects the true mountain profile quite well, and overestimates the
muon fluxes from the West ($\phi \sim \pi$) where the mountain meets
a plateau.} with height 1.3 km and radius of base of 2.1 km. Cosmic
muons produced on the surface were propagated (based on their energy
and direction) to the detector, using the energy-loss formula of
Eqs.~\eqref{mu_loss1},~\eqref{mu-loss2}. These muons were then propagated
inside the detector assuming no magnetic field, for simplicity, and the
corresponding hits/tracks analyzed.  Since the fluxes are very large,
about 4000 events per hour, it was feasible to only generate events for
73 days, about 1/5th of a year. Of the 6527940 cosmic muon events with
energy $1 \le E_\mu \le 300$ GeV, 264803 events did not satisfy the criteria of
hits in at least 3 layers and were eliminated. A further 248059 events
did not pass the selection criteria of $N_{hits}/\cos\theta \ge 15$. The
remaining events were multiplied by 50 to obtain events for 10 years
and sorted into three sets.

The first set corresponded to those events with $15 \le
N_{hits}/\cos\theta < 18$, in which the time difference between the first
and last hits is $\Delta T = 5$--6 ns.  The second set corresponded to
those events with $18 \le N_{hits}/\cos\theta < 21$, in which the time
difference between the first and last hits is $\Delta T = 6$--7 ns. The
final set corresponds to those events which have $N_{hits}/\cos\theta \ge 21$,
in which the time difference between the first and last hits is $\Delta
T > 7$ ns. The total events in each set is given in Table
\ref{tab:cosmic}. The direction mis-id fraction of events can then be
estimated using the probabilistic approach: one in $5.73 \times 10^{-7}$ events with $\Delta T = 5$--6 ns (5--6$\sigma$ deviation in
timing), one in $1.97 \times 10^{-9}$ events with $\Delta T = 6$--7 ns
(6--7$\sigma$ deviation), and none for events having greater than
$\Delta T > 7$ ns. It can be seen from Table \ref{tab:cosmic} that
just 13 cosmic ray background events can be expected in 10 years. We
therefore ignore this background as well.

\begin{table}[htp]
  \centering
  \caption{Cosmic ray events that pass various selection criteria and
the number of these that would be mis-identified in up/down direction
in 10 years. See text for details.}
\label{tab:cosmic}
\begin{tabular}{lrrr} \hline
Event Set & Events in & Probability of & Direction \\
$\Delta T$ ns & 10 years & direction mis-id & Mis-id'ed Events \\ \hline
5--6 & 22349400 & $5.73 \times 10^{-7}$ & 12.8 \\
6--7 & 4558600 & $1.97 \times 10^{-9}$ & $9 \times 10^{-3}$ \\
$\ge 7$ & 291798500 & 0 & 0 \\ \hline
\end{tabular}

\end{table}

Recently, the INO collaboration has measured \cite{John:2022fuy} the direction
mis-identified events in the cosmic ray muon sample detected in the
prototype mini-ICAL detector functioning in Madurai, South India. The
mini-ICAL detector is a $4\times 4$ m$^2$ scaled prototype of ICAL,
about 1 m high, with 11 iron layers, with RPCs populated only in the
central $2\times 2 \times 1$ m$^3$ volume. Hence the muons detected are
purely cosmic ray muons and are all expected to reconstruct in the
downward direction. While the paper includes a detailed analysis of
improving time and position resolutions of the RPC detectors, the
direction mis-identification fraction has also been studied by them for
various selection criteria. The results are presented for tracks with
hits in different number of layers ($N_{hits}=7$--10), with no selection
criteria analogous to the $N_{hits}/\cos\theta$ used here. However,
since the cosmic muons peak around $\theta \sim 30^\circ$ ($1/\cos\theta
= 1.15$), we can roughly compare these results with our probabilistic
approach by assuming $N_{hits}/\langle\cos\theta\rangle \sim
N_{hits}$. Using the argument above, we then have the statistical
probability of direction mis-identification with $N_{hits} = 7,8,9,10$
to be 0.020\%, 0.008\%, 0.003\%, and 0.001\% respectively. The measured
results with reasonable selection criteria were found to be 0.140\%,
0.014\%, 0.0035\% and 0.0014\% respectively, the latter with larger
errors; see Ref.~\cite{John:2022fuy} (and figure~22 therein) for more details. This
agrees with the probabilistic estimates, especially for larger $N_{hits}$
and thus validates our estimates.

\subsection{Other backgrounds}

Finally, it is possible that down-going cosmic ray muons that do not
interact in the detector interact with the rock below the detector to
give up-going neutrons and pions that are detected in ICAL. A study by the
MACRO collaboration \cite{Macro} has found that the background from such
events is about 1\% with a flat zenith angle distribution. Although the
actual number of events depends on the details of the detector sensitivity
and depth at which it is located, this gives a ball-park estimate since
the sizes of MACRO and ICAL are commensurate. Such hadrons generally
shower and will be rejected by the track-finding algorithm that fits
muon tracks. Hence it is expected that these events will not constitute
a significant background to the rock muon events.

In summary, there are different backgrounds to the study of rock muon
events, and they can be reduced by a judicious choice of selection
criteria. The background due to CC atmospheric neutrino events from the
front and back of the detector forms the largest background. However,
in the analysis that follows, we have assumed that the backgrounds can
be kept under check and have not included the impact of these on our
results. There is certainly room for improvement here but this is beyond
the scope of this simplistic analysis. We discuss the physics analysis of upward-going muons in the next section.

\section{Physics Analysis of Upward-going Muons}
\label{phy_an}

\subsection{Event Generation}
\label{evt_gen}

We have used the neutrino event generator NUANCE (version 3.5)
~\cite{nuance} to generate (unoscillated) events corresponding to an exposure of 51 kton
$\times$ 200 years (i.e., 200 years' at ICAL) of unoscillated upward-going
muons in the energy range 0.8--200 GeV. The atmospheric neutrino
fluxes as provided by Honda et al.~\cite{honda} at the Super Kamiokande
experiment~\cite{sk} were used. The ICAL detector specifications were
defined inside NUANCE. Dimensions of the detector were chosen such that no
events were generated inside the detector i.e., only its external geometry
was used. The actual material in which interactions occur is rock,
whose density was taken to be 2.65 gm/cm$^{3}$. To cut off cosmic ray
backgrounds, only events arriving with zenith angles\footnote{Note that rock muons are produced
in the rock surrounding the detector so that they can enter the
detector from above, below, or any of the four sides. The ones entering
from above are lost in the huge cosmic ray muon background and cannot
be detected. Hence here we only detect rock muons entering from the
bottom and sides of the detector.} $0^{\circ} < \theta < 81^{\circ}$
($\cos\theta > 0.156$), were selected~\cite{kan}. The NUANCE generator
itself propagates the muons produced in the CC interaction to the closest
surface of the ICAL detector by using appropriate energy loss formulae
as discussed earlier in Eqs.~\eqref{mu_loss1} and \eqref{mu-loss2}. The
hadrons are absorbed in the rock and so only muons enter the detector.
The NUANCE generator gives information not only of the initial vertex
of the CC neutrino interaction but also the energy and direction of the
initial neutrino and the produced muon. The information on the neutrino
energy and zenith angle $(E_\nu, \cos\theta_\nu)$ is saved for later
use to oscillate the data set according to the neutrino oscillations. 

Two data sets were generated: set I with ``normal'' fluxes, and set
II with ``swapped'' fluxes where the muon and electron neutrino fluxes
were interchanged. This was used to enable incorporation of oscillations
later. Note that the energy and direction of the neutrino initiating the
event is also available in NUANCE. These values are required to generate
the correct neutrino oscillation and survival probabilities, as
described later. The information on the vertex and energy and direction
of the produced muons is used to propagate the muons to the detector
surface, as described earlier. Only those events falling within the
aperture of the detector are retained; hence generating rock muons is a
time-consuming process in contrast to the ``standard'' analysis where
the CC interaction occurs inside the detector itself.

The number of muon events from the two channels are generically given by,
\begin{eqnarray} 
\frac{{\rm d}^2N^-_{\mu\mu}}{{\rm d}E_\mu^t {\rm d} \cos\theta^t_\mu}   
	& = &  T \times N_D \int_0^\infty {\rm d}E_\nu
	\int_{-1}^{1} {\rm d} \cos\theta_\nu
	\int_{0}^{2\pi} {\rm d} \phi_\nu \times \nonumber \\
 & & \frac{{\rm d}^3\Phi_{\mu}}{{\rm d}E_{\nu}{\rm d}\cos\theta_{\nu}
 {\rm d}\phi_\nu} \times \frac{{\rm d}^2 \sigma_{\nu_{\mu}}}
	{{\rm d}E_\mu^t {\rm d} \cos\theta^t_\mu}~,  \\
\frac{{\rm d}^2N^-_{e\mu}}{{\rm d}E_\mu^t {\rm d} \cos\theta^t_\mu}    
	& = &  T \times N_D \int_0^\infty {\rm d}E_\nu
	\int_{-1}^{1} {\rm d} \cos\theta_\nu
	\int_{0}^{2\pi} {\rm d} \phi_\nu \times \nonumber \\
 & & \frac{d^{3}\Phi_{e}}{dE_{\nu}d\cos\theta_{\nu} {\rm d} \phi_\nu} \times 
\frac{{\rm d}^2 \sigma_{\nu_{\mu}}} {{\rm d}E_\mu^t {\rm d}
\cos\theta^t_\mu}~.
\label{eqntd}
\end{eqnarray}
The label `{\it t}' refers to the true values of the muon energy and
zenith angle; $T$ is the exposure time and $N_D$ is the total number of target
nucleons in the rock, which is assumed to surround the detector to an
infinite distance in all directions. The produced muons
are then propagated \cite{nuance} in the rock with suitable energy loss
until they reach the detector. In order to speed up
the events generation, two quantities can be specified: the minimum
energy of the muon when it reaches the closest face of the detector, and
the maximum zenith angle $\theta$; the latter prevents the generation of the
uninteresting horizontal and down-going events. Here $\Phi_{\mu}$ and
$\Phi_{e}$ are the atmospheric fluxes of ${\nu_{\mu}}$ and ${\nu_{e}}$
respectively and similar equations hold for $\mu^+$ produced from CC
interactions of muon anti-neutrinos.

Instead of being passed through GEANT, the muon energy and angle
corresponding to these NUANCE events were then smeared according
to the resolutions and reconstruction efficiencies obtained in
section~\ref{res_eff} so that the event was binned according to its
smeared/observed energy and angle values. Neutrino oscillation is then
applied as follows.

The set of input neutrino oscillation parameters used in the analysis
are listed in Table~\ref{table1}. Since the analysis is not sensitive
to the 1--2 parameters, these were kept fixed throughout. In addition,
these events are not sensitive to the CP phase, which was also kept
fixed. The normal ordering was assumed to be the true one as well.

\begin{table}[tbp]
  \centering
  \caption{Values of neutrino oscillation parameters used
in this study~\cite{lsmohan}. The second column shows the central values
of the oscillation parameters while the third column shows the 3$\sigma$
ranges of the parameters. Normal hierarchy (NH) is assumed throughout.}
\label{table1} 
\begin{tabular}{|c|c|c|}
\hline 
Parameter & Central/input values & 3$\sigma$ ranges \\ 
\hline
$\Delta m^{2}_{21}$ (eV$^{2}$) & 7.5 $\times$ $10^{-5}$ & fixed \\
$\Delta m^{2}_{32}$ (eV$^{2}$) & 2.4 $\times$ $10^{-3}$ (NH) & [2.1, 2.6] $\times$ $10^{-3}$ (NH) \\
$\sin^{2}\theta_{12}$ & 0.304 & fixed \\
$\sin^{2}\theta_{23}$ & 0.5 & [0.360, 0.659] \\
$\sin^{2}\theta_{13}$ & 0.022 & [0.018, 0.028] \\
$\delta_{CP}$ ($^{0}$) & 0 & fixed \\
\hline
\end{tabular}

\end{table}

Due to the presence of both $\nu_{e}$ and $\nu_{\mu}$ atmospheric fluxes,
there are contributions from two channels, viz., the survived $\nu_{\mu}$
neutrinos, determined by $P_{\mu\mu}$ and the oscillated $\nu_e$
neutrinos, determined by $P_{e\mu}$, to the muon events in ICAL. Hence
each event in set I is oscillated according to $P_{\mu\mu} (E_\nu,
\cos\theta_\nu)$ as determined by the oscillation parameters,
while each event in set II is oscillated according to $P_{e\mu} (E_\nu,
\cos\theta_\nu)$.

To implement oscillations we have used a re-weighting algorithm as
follows. We generated a uniform random number $r$ between 0 and 1; if
$P_{\mu \mu} > r$, then the event survives oscillations and is binned
appropriately; similarly, if $P_{e\mu} > r$ we considered the swapped
event to contribute as an oscillated $\nu_e\to \nu_\mu$ event. Events from
both channels were added to get the total $\mu^{-}$ events. Symbolically,
we have:
\begin{eqnarray}
N_\mu^- = P_{\mu\mu} N_{\mu\mu}^- + P_{e\mu} N_{e\mu}^- ~.
\label{eq:Ntot}
\end{eqnarray}
A similar procedure was applied to get $\mu^{+}$ events from $\overline
\nu$, with the corresponding anti-neutrino survival/oscillation
probabilities.

\begin{figure*}[btp]
  \centering

\includegraphics[width=0.4\textwidth,height=0.25\textwidth]{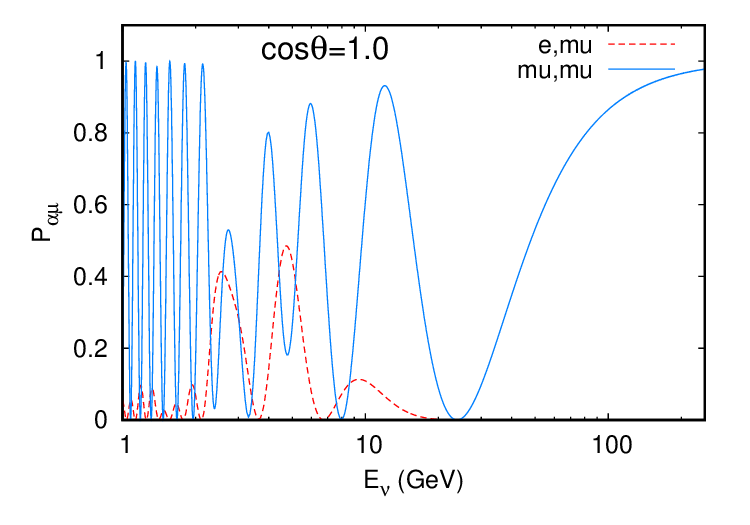}
\includegraphics[width=0.4\textwidth,height=0.25\textwidth]{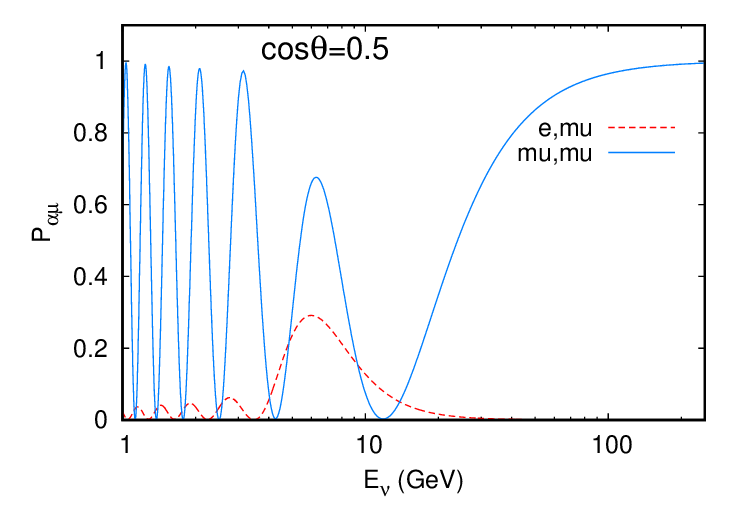}
\includegraphics[width=0.4\textwidth,height=0.25\textwidth]{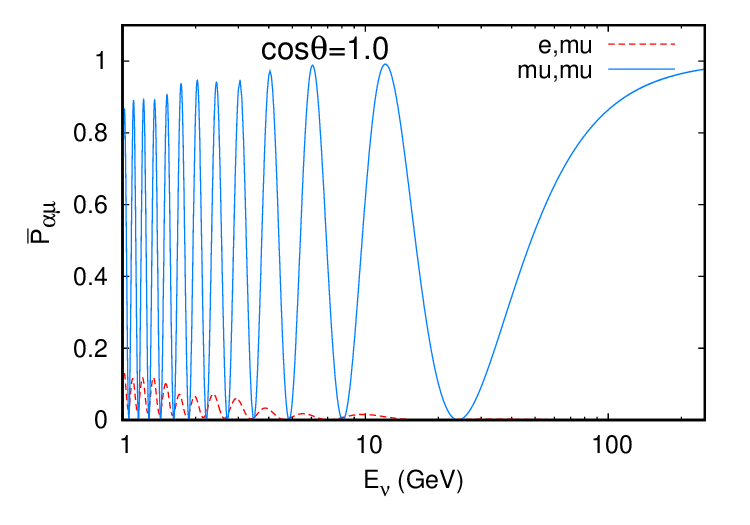}
\includegraphics[width=0.4\textwidth,height=0.25\textwidth]{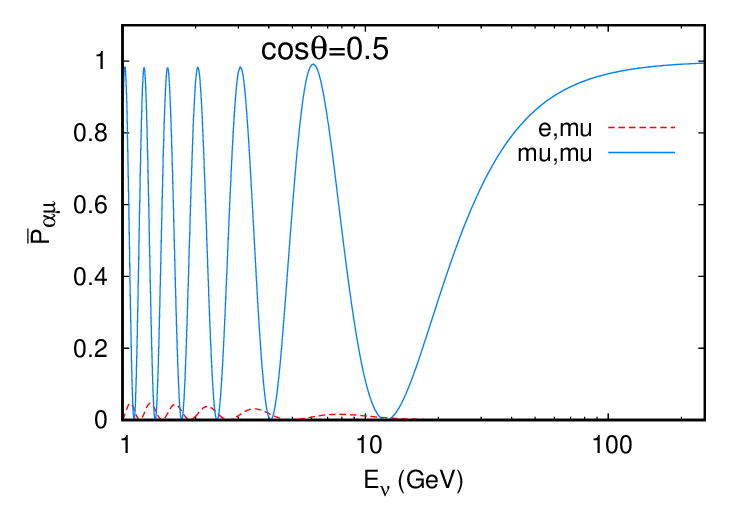}
\caption{The survival and oscillation probabilities, $P_{\alpha\mu}$ and
$\overline{P}_{\alpha\mu}$ for $\alpha=e,\mu$ as a function of the
neutrino energy for two different zenith angles, $\cos\theta = 0.5, 1.0$.}
\label{fig:oscprob}
\end{figure*}

Note that $P_{e\mu} \ll P_{\mu\mu}$, as can be seen from figure~\ref{fig:oscprob} where the relevant survival and oscillation
probabilities for both neutrinos and anti-neutrinos have been plotted
for two different values of the zenith angle, $\cos\theta = 0.5,
1.0$~\cite{kan1}.

The oscillated data was binned into bins of observed/ smeared muon
energy and $\cos\theta$ bins. We have used two schemes of binning in
this paper. Firstly, since there were substantial events with energy
$E_\mu \gtrsim 100$ GeV, we took 25 bins of smeared energy as given in
Table~\ref{E_cos3}; we refer to this as exponential binning. The energy
bins were optimized such as to obtain reasonable number of events
in each bin. Secondly, we took linear energy bins with the width of
1 GeV from 1--45 GeV for comparison with exponential binning scheme
to check sensitivity to oscillation parameters. The data sample has
a proportionately larger component of higher energy events which were
not sensitive to oscillations, so finer bins were used at lower energy.
In each case, the data was divided into seven bins of $\cos\theta$ from
0 to 1; 6 uniform ones of width 0.15, with the last bin from 0.9--1.0.

\begin{table}[tbp]
  \centering
  \caption{\label{E_cos3} Choice of observed/smeared energy bins of muons
for the case of exponential binning.}
\begin{tabular}{|c|c|c|}
\hline 
   Energy range (GeV) & Bin width (GeV) & No. of bins \\
  \hline
	1--9 & 1 & 8 \\ 
	9--17 & 2 & 4 \\ 
	17--20 & 3 & 1 \\ 
	20--40 & 5 & 4 \\
	40--80 & 10 & 4 \\ 
	80--100 & 20 & 1 \\ 
	100--200 & 50 & 2 \\ 
	200--256 & 56 & 1 \\ \hline
\end{tabular}
\end{table}

The number of $\mu^{\pm}$ events observed in a given bin $(i,j)$ of
observed $(E_{\mu}^i, \cos\theta_{\mu}^j)$ after oscillations, and on
including the detector response (smearing of muon energy and angle as
well as including reconstruction and cid efficiencies) are then given by,

 \begin{eqnarray}
 N_{\mu}^-(i,j) & = & \epsilon_{R} \times [ \epsilon^{-}_{C}
 \times N_{\mu}^{-} (E_{\mu}^{i}, \cos\theta_{\mu}^{j}) +
 (1 - \epsilon^{+}_{C}) \times N_{\mu}^{+} (E_{\mu}^{i},\cos\theta_{\mu}^{j}) ],
 \label{nc}
\end{eqnarray}

\begin{eqnarray}
 N_{\mu}^+(i,j) & = & \epsilon_{R} \times [ \epsilon^{+}_{C}
 \times N_{\mu}^{+} (E_{\mu}^{i}, \cos\theta_{\mu}^{j}) +
 (1 - \epsilon^{-}_{C}) \times N_{\mu}^{+} (E_{\mu}^{i},\cos\theta_{\mu}^{j}) ]
\label{nc1}
\end{eqnarray}

where $N_{\mu}^{\pm}$ are the total number of $\mu^{\pm}$ events
in the $(i,j)^{th}$ bin after detector smearing and oscillations,
$\epsilon_{C}^{\pm}$ is the cid efficiency (here $\epsilon_{C}^{+} =
\epsilon_{C}^{-}$), and $\epsilon_{R}$ is the reconstruction efficiency
(which is the same for $\mu^\pm$). Note that $\epsilon_{R}$ and
$\epsilon_{C}$ have been determined from simulations as functions of
the true energy and angle ($E_{\mu}^{t}, \cos\theta^{t}$) of the muons,
while $N_{\mu}^{\pm}$ ($E_{\mu}, \cos\theta_{\mu}$) refer to the smeared
(or, in the actual experiment, observed) values for the muons. The second
term in Eqs.~\eqref{nc} and ~\eqref{nc1} is due to the charge mis-identification, and the oscillated events themselves can be obtained from the expressions given in Eqs.~\eqref{eqntd} and \eqref{eq:Ntot}. The events oscillated according
to the input parameters mentioned earlier have been scaled down to 4.5 or
10 years and labelled as ``data''. The same set was scaled but oscillated
according to an arbitrary set of oscillation parameters and referred to as
``theory'' in this simulation analysis.

\begin{figure*}[bhp]
  \centering
  \includegraphics[width=0.32\textwidth,height=0.23\textwidth]{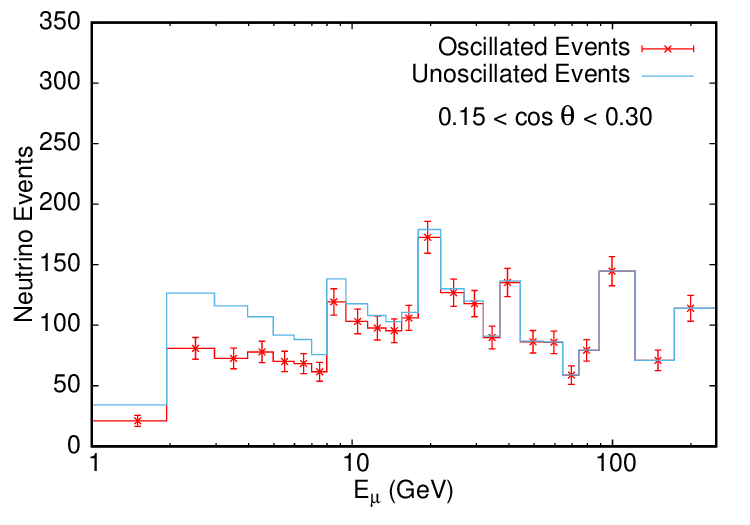}
\includegraphics[width=0.32\textwidth,height=0.23\textwidth]{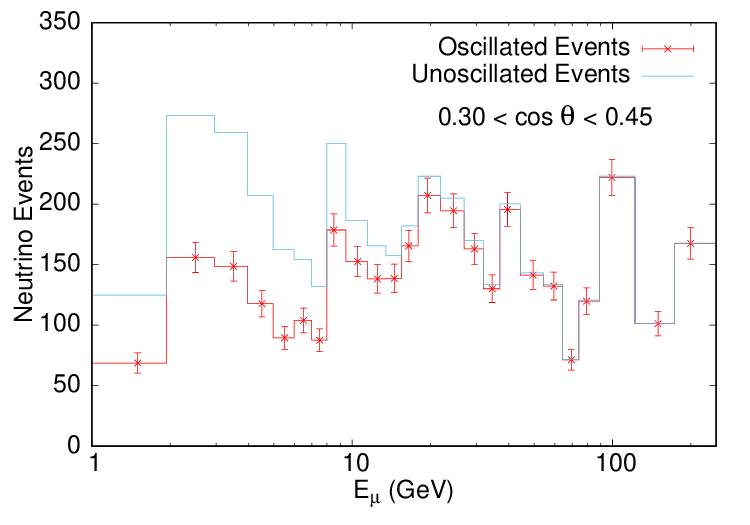}
\includegraphics[width=0.32\textwidth,height=0.23\textwidth]{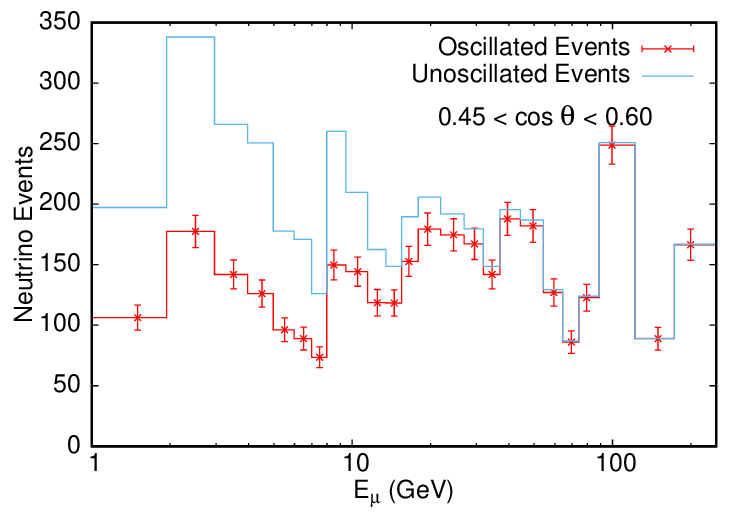}
\includegraphics[width=0.32\textwidth,height=0.23\textwidth]{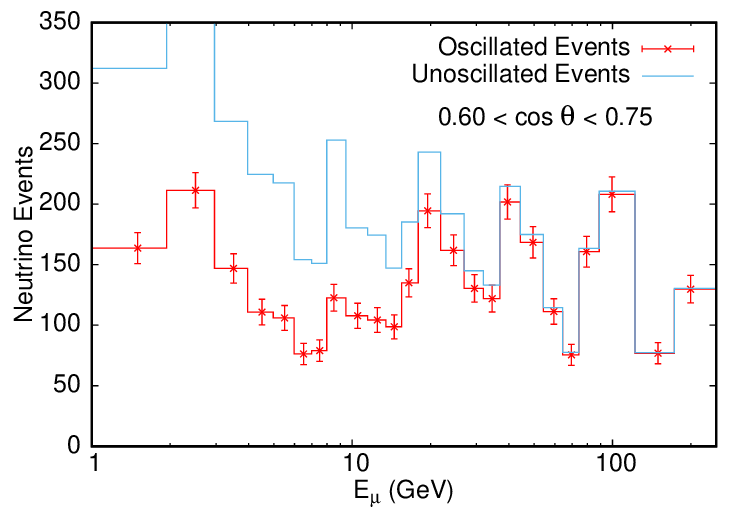}
\includegraphics[width=0.32\textwidth,height=0.23\textwidth]{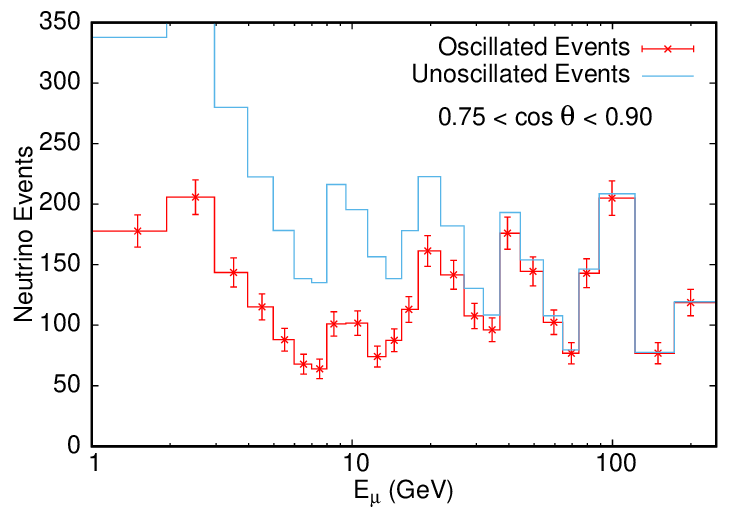}
\includegraphics[width=0.32\textwidth,height=0.23\textwidth]{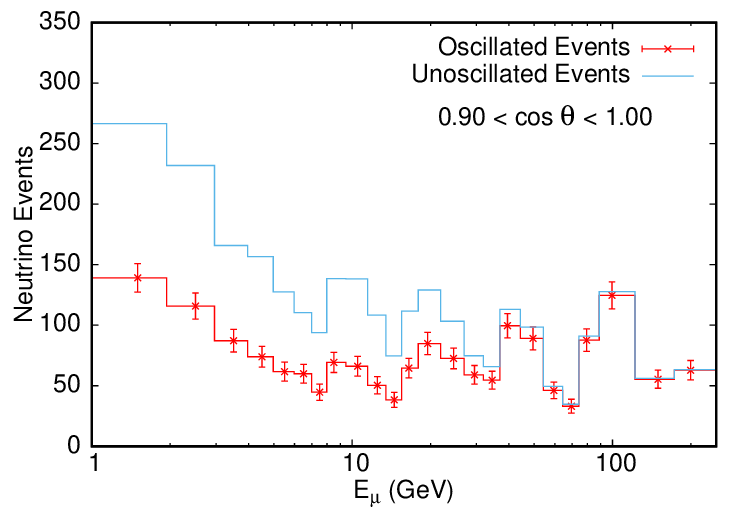}

\caption{Rock muon ($\mu^-$) events as a function of observed
energy $E_\mu$ for six of the seven $\cos\theta$ bins, excluding the most
horizontal one, with $\sqrt{N}$ errors. The central values of oscillation
parameters as given in Table~\ref{table1} have been used. The unoscillated
events are shown in blue.}
\label{fig:energy_mum}
\end{figure*}

\begin{figure*}[htp]
  \centering
  \includegraphics[width=0.32\textwidth,height=0.23\textwidth]{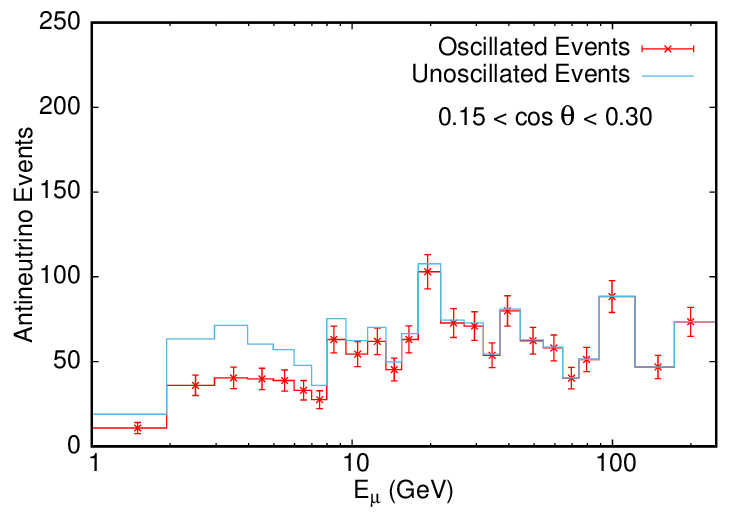}
\includegraphics[width=0.32\textwidth,height=0.23\textwidth]{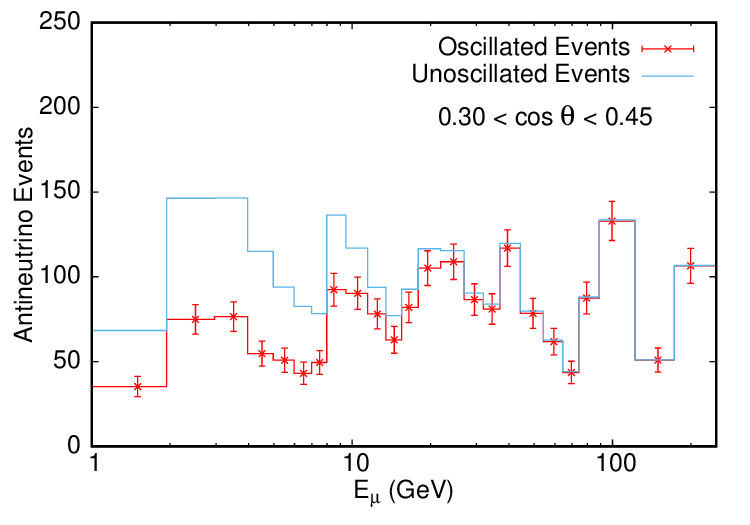}
\includegraphics[width=0.32\textwidth,height=0.23\textwidth]{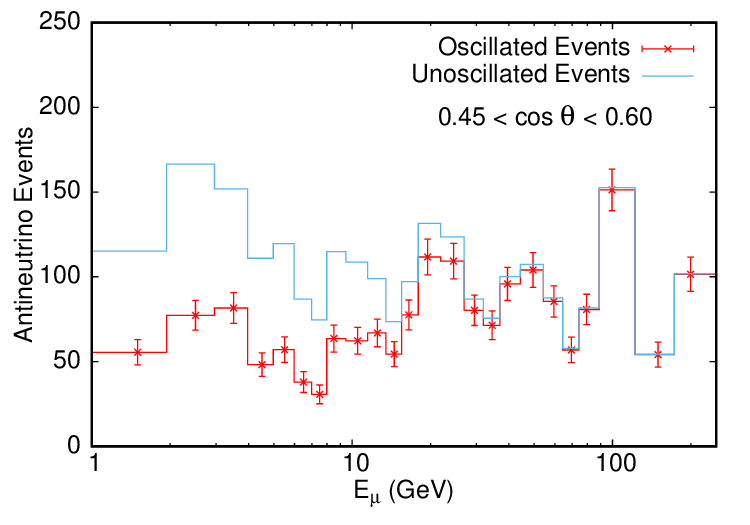}
\includegraphics[width=0.32\textwidth,height=0.23\textwidth]{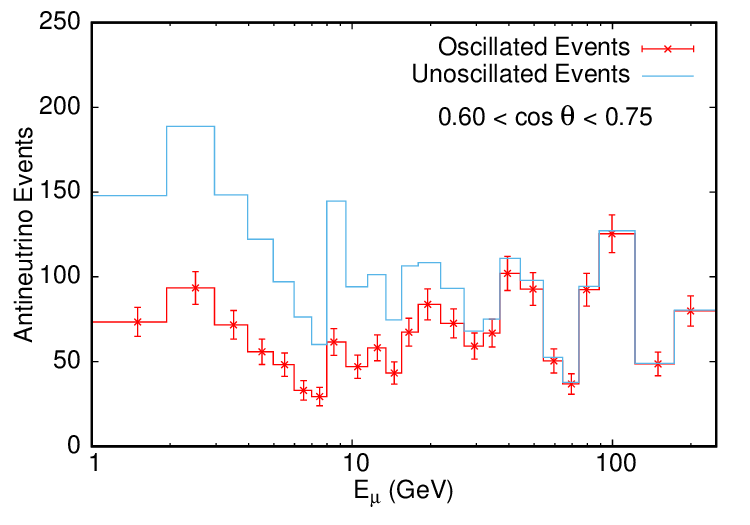}
\includegraphics[width=0.32\textwidth,height=0.23\textwidth]{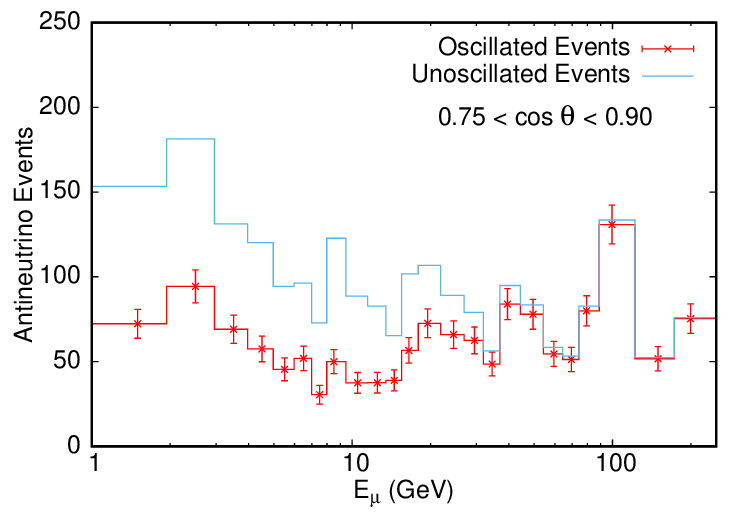}
\includegraphics[width=0.32\textwidth,height=0.23\textwidth]{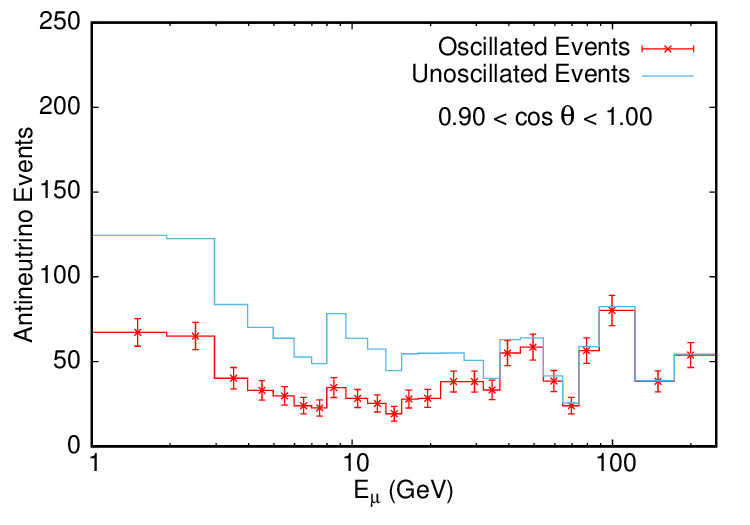}

\caption{As in figure~\ref{fig:energy_mum} for rock muon ($\mu^+$) events.}
\label{fig:energy_mup}
\end{figure*}

The energy distribution of muons ($E_{\mu}$) for different $\cos\theta$
bins is shown in figures~\ref{fig:energy_mum} and \ref{fig:energy_mup} for
muon and anti-muon events. The events fall with increasing
energy. Note that the ``oscillating'' nature of the distribution is due
to the fact that the bin sizes are not uniform; see Table~\ref{E_cos3}
and associated discussion.

Events at higher energies beyond about 20 GeV are not sensitive to
oscillations; however, they cannot be neglected despite being small
in number as they also contribute to the statistics and help in
flux normalization since the higher energy cross sections are known
better. There are just a few events in the first $\cos\theta$ bin i.e.,
0.0--0.15, reflecting the poorer resolutions and reconstruction
efficiencies at large angles, as seen in figures~\ref{fig:eff}
and~\ref{fig:resol}.

\subsection{Best fit Analysis}
\label{chi_an}

A $\chi^{2}$ analysis has been done, taking into account systematic
uncertainties through the pulls method \cite{maltoni}. The $\chi^{2}$
analysis uses the ``data'' binned in the observed momentum and zenith
angle of the muons, where we have included only events with reconstructed
muon energy, $E_\mu < 200$ GeV. In particular, for the physics analysis,
we have used the energy range $1 \le E_\mu \le 200$ GeV for bottom
events, $4 \le E_\mu \le 200$ GeV for the side events (due to the poor
reconstruction we did not take into account the higher energy events),
and have only considered events where the reconstructed muon angle,
$\cos\theta_\mu > 0.2$. The number of events in 10 years exposure at
ICAL is given in Table~\ref{tab:events}.

\begin{table*}%[btp]
  \centering
  \small
  \caption{\label{tab:events}The number of oscillated and unoscillated
$\mu^{-}$ and $\mu^+$ events in 10 years in bins of reconstructed
$E_{\mu}$ (GeV) and $\cos\theta_{\mu}$ at the best fit value of the
neutrino oscillation parameters given in Table~\ref{table1}. Note that
finite detector resolutions and efficiencies have been folded into
the results.}
\begin{tabular}{|r|r|r|r|r|r|r|r|} \hline
$\cos\theta_{min}$ & $\cos\theta_{max}$ & $E^{\mu}_{min}$ & $E^{\mu}_{max}$ & 
\multicolumn{2}{|c|}{Oscillated Events} &
\multicolumn{2}{|c|}{Unoscillated Events}  \\ 
& & (GeV) & (GeV) & ~~~~$\mu^-$  & $\mu^+$ & ~~~~$\mu^-$ & $\mu^+$ \\ \hline
    0.15  &   0.30  &    1.0  &    9.0 &  23  & 11 &  32  & 18 \\
    0.15  &   0.30  &    9.0  &   17.0 &  20  & 11 &  23  & 12 \\
    0.15  &   0.30  &   17.0  &   20.0 &   5  &  3 &   5  &  3 \\
    0.15  &   0.30  &   20.0  &   40.0 &  26  & 15 &  26  & 16 \\
    0.15  &   0.30  &   40.0  &   80.0 &  19  & 12 &  20  & 12 \\
    0.15  &   0.30  &   80.0  &  100.0 &   3  &  2 &   3  &  2 \\
    0.15  &   0.30  &  100.0  &  200.0 &  10  &  6 &  10  &  6 \\ \hline
    0.30  &   0.45  &    1.0  &    9.0 &  41  & 20 &  70  & 39 \\
    0.30  &   0.45  &    9.0  &   17.0 &  30  & 16 &  38  & 21 \\
    0.30  &   0.45  &   17.0  &   20.0 &   8  &  4 &   9  &  4 \\
    0.30  &   0.45  &   20.0  &   40.0 &  36  & 19 &  38  & 20 \\
    0.30  &   0.45  &   40.0  &   80.0 &  29  & 16 &  30  & 17 \\
    0.30  &   0.45  &   80.0  &  100.0 &   5  &  4 &   6  &  4 \\
    0.30  &   0.45  &  100.0  &  200.0 &  16  &  9 &  16  &  9 \\ \hline
    0.45  &   0.60  &    1.0  &    9.0 &  46  & 22 &  86  & 47 \\
    0.45  &   0.60  &    9.0  &   17.0 &  26  & 12 &  39  & 19 \\
    0.45  &   0.60  &   17.0  &   20.0 &   7  &  3 &   9  &  4 \\
    0.45  &   0.60  &   20.0  &   40.0 &  33  & 18 &  38  & 21 \\
    0.45  &   0.60  &   40.0  &   80.0 &  31  & 17 &  33  & 18 \\
    0.45  &   0.60  &   80.0  &  100.0 &   6  &  4 &   6  &  4 \\
    0.45  &   0.60  &  100.0  &  200.0 &  16  & 10 &  16  & 10 \\ \hline
    0.60  &   0.75  &    1.0  &    9.0 &  54  & 24 & 104  & 50 \\
    0.60  &   0.75  &    9.0  &   17.0 &  21  & 10 &  37  & 20 \\
    0.60  &   0.75  &   17.0  &   20.0 &   6  &  3 &   9  &  5 \\
    0.60  &   0.75  &   20.0  &   40.0 &  31  & 14 &  38  & 18 \\
    0.60  &   0.75  &   40.0  &   80.0 &  30  & 15 &  31  & 16 \\
    0.60  &   0.75  &   80.0  &  100.0 &   8  &  4 &   8  &  4 \\
    0.60  &   0.75  &  100.0  &  200.0 &  14  &  8 &  14  &  8 \\ \hline
    0.75  &   0.90  &    1.0  &    9.0 &  56  & 26 & 109  & 54 \\
    0.75  &   0.90  &    9.0  &   17.0 &  18  &  8 &  35  & 17 \\
    0.75  &   0.90  &   17.0  &   20.0 &   5  &  2 &   8  &  5 \\
    0.75  &   0.90  &   20.0  &   40.0 &  26  & 12 &  35  & 18 \\
    0.75  &   0.90  &   40.0  &   80.0 &  25  & 13 &  28  & 14 \\
    0.75  &   0.90  &   80.0  &  100.0 &   7  &  3 &   7  &  4 \\
    0.75  &   0.90  &  100.0  &  200.0 &  14  &  9 &  14  &  9 \\ \hline
    0.90  &   1.00  &    1.0  &    9.0 &  39  & 19 &  78  & 38 \\
    0.90  &   1.00  &    9.0  &   17.0 &  11  &  5 &  22  & 12 \\
    0.90  &   1.00  &   17.0  &   20.0 &   3  &  1 &   5  &  2 \\
    0.90  &   1.00  &   20.0  &   40.0 &  14  &  6 &  20  & 10 \\
    0.90  &   1.00  &   40.0  &   80.0 &  14  &  9 &  16  & 10 \\
    0.90  &   1.00  &   80.0  &  100.0 &   4  &  2 &   4  &  2 \\
    0.90  &   1.00  &  100.0  &  200.0 &   9  &  5 &   9  &  6 \\ \hline 
\end{tabular}

\end{table*}

Five different sets of systematic uncertainties~\cite{maltoni}
were considered for our analysis as given in~\cite{lakshmi}: a flux
normalization error of $20\%$, 10\% error on cross-sections, 5\% error
on zenith angle dependence of flux, and an energy dependent tilt error,
which is described as follows. The event spectrum have been calculated
with the predicted atmospheric neutrino fluxes and with the flux spectrum
shifted as,
\begin{eqnarray}
\Phi_{\delta}(E) = \Phi_{0}(E) \Bigg(\frac{E}{E_{0}}\Bigg)^{\!\delta}
\simeq \Phi_{0}(E) \Bigg(1 + \delta \ln \frac{E}{E_{0}}\Bigg) ~.
\label{eq:tilt}
\end{eqnarray}
The different parameters are, $E_{0}$ = 2 GeV, $\delta$ = 1$\sigma$
systematic tilt error, which was taken as 5\%. In addition, an
overall systematic of 5\% was taken to account for uncertainties
such as those arising from the reconstruction of the muon energy and
direction, due to uncertainties in the magnetic field used in the
Kalman filter~\cite{Kolahal}. We list these below. Two different analyses were performed,
one where the $\mu^{+}$ and $\mu^{-}$ events were separately considered,
and the other where they were combined into the same bins (charge-blind
analysis). The former uses 10 pulls, 5 for each charge sign, while the
latter uses 5 (common) pulls. Since the events in each bin are small,
we use the Poissonian definition of $\chi^{2}$~\cite{physics}. The
Poissonian definitions for 10 pulls is given by:
\begin{eqnarray}
\chi^{2}_{\pm}  & = & \sum\limits_{i=1,j=1} \Bigg[2\Bigg(
N^{\pm, \rm th}_\mu(i,j) - N^{\pm, \rm obs}_\mu(i,j)  \Bigg) \nonumber \\
 & & ~~~~~ - 2\,N^{\pm, \rm obs}_\mu(i,j) 
\times \ln \Bigg(\frac{N^{\pm, \rm th}_\mu(i,j)}{N^{\pm, \rm obs}_\mu(i,j)}
\Bigg)\Bigg],  \\ 
\chi^{2} & = & \chi^{2}_{-} + \chi^{2}_{+} + \sum\limits_{k=1}^{5}
\left(\left(\xi_k^-\right)^2 +
\left(\xi_k^+\right)^2~\right).
\label{chi1}
\label{poisson}
\end{eqnarray} 
Here, 
\begin{eqnarray}
N^{\pm, \rm th}_\mu(i,j) = N^{\pm}_\mu(i,j) 
	\left(1 + \sum\limits_{k=1}^{5} \pi^{k}_{ij} \xi_{k}^\pm\right)~,
\label{chi2}
\end{eqnarray} 
where $N_{ij}^{th}$ , $N_{ij}^{obs}$ are the theoretically predicted
and ``observed'' data in given bins of ($E_{\mu}$, $\cos\theta$),
and $N^\pm_\mu(i,j)$ are the number of events without the systematic
uncertainties defined in Eqs.~\eqref{nc} and ~\eqref{nc1}. Here $\pi^{k}_{ij}$
are the (common) systematic errors for both $\mu^-$ and $\mu^+$ events,
given in Table~\ref{tab:pulls}, and $\xi_{k}^\pm$
are the pull variables which are solved for by minimising the $\chi^2$
for each set of oscillation parameters. Hence the $\chi^{2}_{\pm}$ minimization has been done
independently, first over the pulls for a given set of oscillation
parameters, and then over the oscillation parameters themselves. In
the analysis where the charge of the muon was not determined, the total
muon events, $N_\mu = (N^-_\mu + N^+_\mu)$, were binned into the same
observed ($E_{\mu}$, $\cos\theta$) bin and a common pull was applied to
the summed events:
\begin{eqnarray}
\chi^{2}_{sum} & = &  \sum\limits_{i=1,j=1} \Bigg[2\Bigg(
N^{\rm th}_\mu(i,j) - N^{\rm obs}_\mu(i,j)  \Bigg) - \nonumber \\
 & & \qquad 2\,N^{\rm obs}_\mu(i,j)
\ln \Bigg(\frac{N^{\rm th}_\mu(i,j)}{N^{\rm obs}_\mu(i,j)}
\Bigg)\Bigg] + \sum\limits_{k=1}^5 \left(\xi_k\right)^2~,
\label{chi2sum}
\end{eqnarray}
and only 5 pulls were used in the analysis.

\begin{table*}%[tbp]
  \centering
  \caption{\label{tab:pulls} Systematic uncertainties used in our analysis for bins of observed muon energy and angle, $E_i^{obs}$, $\cos\theta_j^{obs}$.}

\begin{tabular}{|c|c|c|}
\hline 
 S. No. &  Systematic errors, $\pi^k_{ij}$ & Value \\
  \hline
 1 & Flux normalization pull, $\pi^1$ & 20 \% \\
 2 & Cross-sections error, $\pi^2$ & 10 \% \\
 3 & Zenith angle error, $\pi^3_j$ & $5\times \cos\theta_j$ \% \\
 4 & Energy dependent tilt error $\pi^4_i$ & Calculated from
Eq.~\ref{eq:tilt} \\
 5 & Reconstruction error $\pi^5$ & 5 \% \\ \hline
\end{tabular}
\end{table*}

Finally, the whole data was marginalized over the 3$\sigma$ ranges
of $\sin^2\theta_{23}$, $\Delta m^{2}_{32}$ and $\theta_{13}$ given
in Table~\ref{table1} and a prior included on $\sin^{2}2\theta_{13}$,
as given by,
\begin{eqnarray}
\chi^{2} = \chi^{2} + \Bigg(\frac{\sin^{2}2\theta_{13}(\hbox{true}) -
\sin^{2}2\theta_{13}}{\sigma_{\sin^{2}2\theta_{13}}} \Bigg)^{2} ~,
\label{th13}
\end{eqnarray}
where $\sigma_{\sin^{2}2\theta_{13}}$ is the 1$\sigma$ error
for the corresponding neutrino parameter which has been taken to be 8\% in
this analysis.

In order to determine $\chi^{2}_{min}$, the minimization of $\chi^{2}$
has been done over all three parameters $\sin^2\theta_{23}$, $\Delta
m^{2}_{32}$ and $\theta_{13}$, keeping the other parameters fixed at
their input values. In order to determine the sensitivity of rock muon
events to a given neutrino oscillation parameter, the change in $\chi^2$,
\begin{equation}
\Delta \chi^2 = \chi^2(\hbox{par})  - \chi^2(\hbox{min}) ~,
\label{eq:delchi}
\end{equation}
is calculated, where the events are generated using the minimum (min)
value of the parameter, and after adding systematic uncertainties and
priors (par). More than one parameter can be changed in the study.

\section{Results} \label{osc_res}

\subsection{Sensitivity to Individual Parameters}
\label{sen_par}

\begin{figure*}[htp]
  \centering
\includegraphics[width=0.49\textwidth]{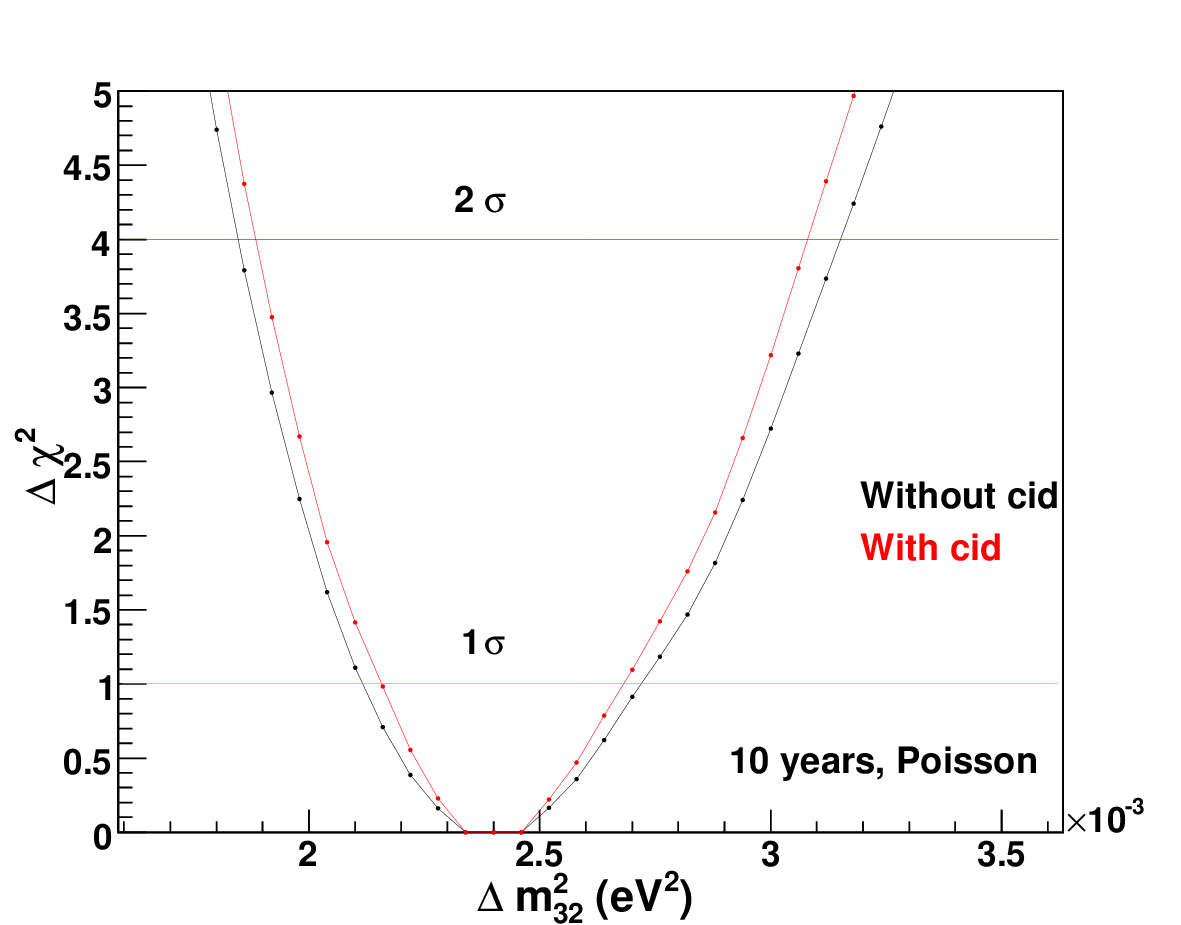}
\includegraphics[width=0.49\textwidth]{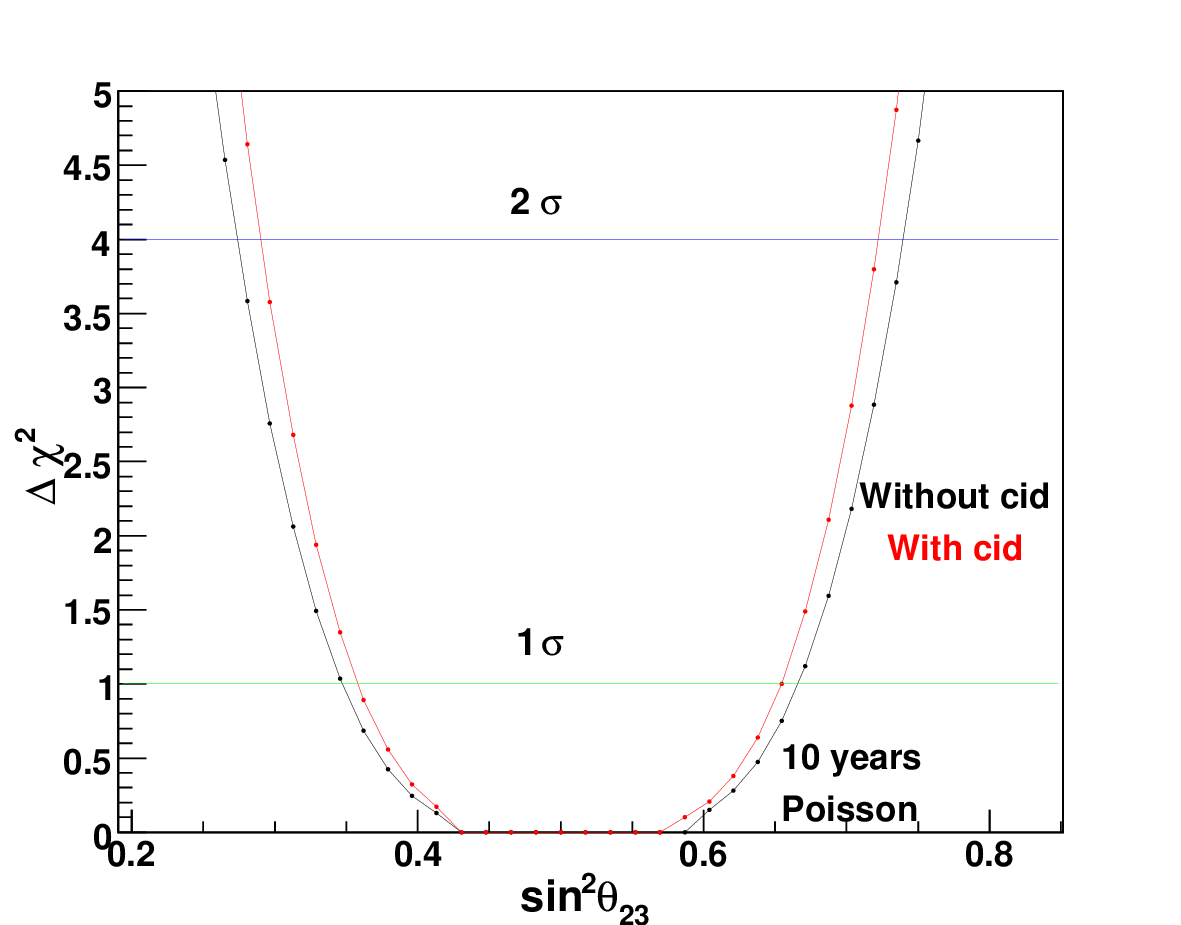}
\caption{A comparison of the sensitivity in $\Delta \chi^2$ with combined
muon events (without cid) (black) with an analysis including muon charge
identification (with cid) with 10 systematic errors (red) as a function
of $\Delta m^{2}_{32}$ (left panel) and $\sin^{2}\theta_{23}$ (right panel) when their input values were taken to be $\Delta m^{2}_{32} (in) = 2.4 \times 10^{-3}$
eV$^2$ and $\sin^2\theta_{23} (in) = 0.50$.}
\label{fig:delm-g5n10}
\end{figure*}

Figure~\ref{fig:delm-g5n10} shows $\Delta\chi^{2}$ as a function of
$\Delta m^{2}_{32}$ using input values of $\Delta m^{2}_{32}$ (true)
= 2.4 $\times$ $10^{-3}$ eV$^{2}$ and $\sin^2\theta_{23}$ = 0.50
with $N_{\mu}^{+}$ and $N_{\mu}^{-}$ events considered together and
separately.  It can be seen that the analysis with charge identification
(cid) efficiency included (that is, separating $N_{\mu}^{+}$ and
$N_{\mu}^{-}$ events) gives a better sensitivity than with combined
events. Figure~\ref{fig:delm-g5n10} also shows a similar plot for
$\Delta\chi^{2}$ as a function of $\sin^{2}\theta_{23}$, with similar
improvement in the cid-dependent analysis. Henceforth, we shall include
charge identification in the analysis. In both cases, the exponential
energy binning scheme of Table~\ref{E_cos3} was used.

\begin{figure*}[htp]
  \centering
  \includegraphics[width=0.49\textwidth]{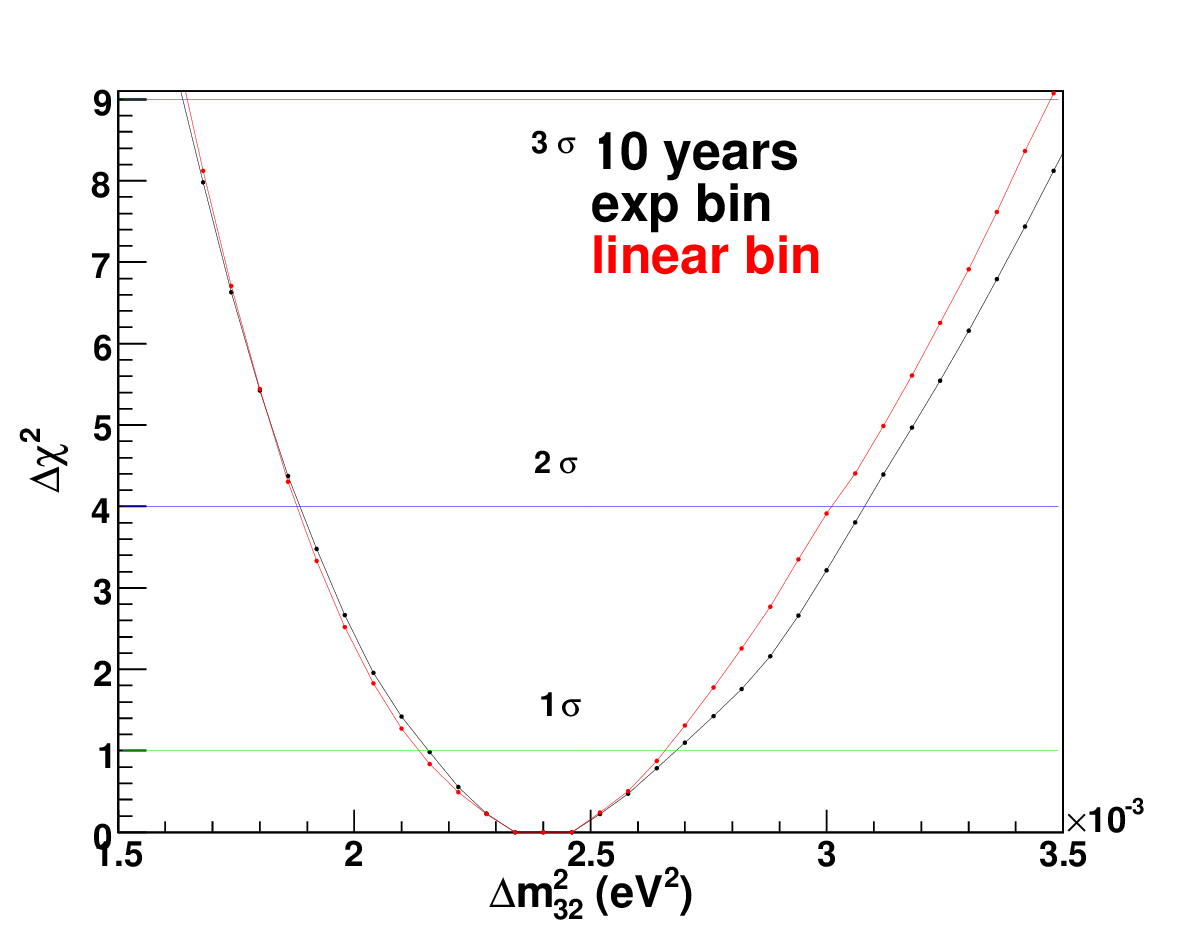}
  \includegraphics[width=0.49\textwidth]{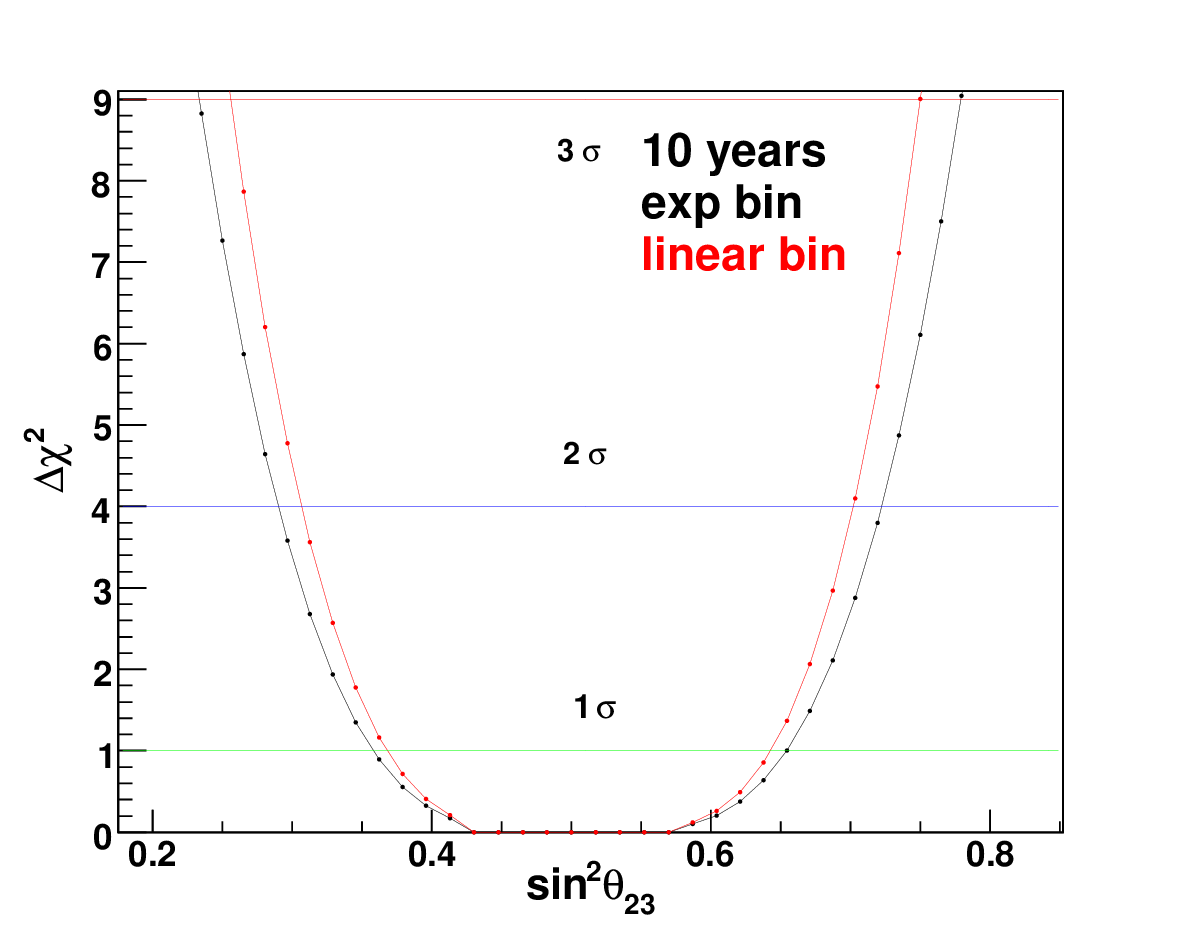}
\caption{A comparison of the sensitivity in $\Delta \chi^2$ with
linear binning (red) and exponential binning (black) as a function of $\Delta m^{2}_{32}$ (left panel) and $\sin^{2}\theta_{23}$ (right panel) when their input
values were taken to be $\Delta m^{2}_{32} (in) = 2.4 \times 10^{-3}$
eV$^2$ and $\sin^2\theta_{23} (in) = 0.50$ and charge identification
has been included.}
\label{fig:del_linear}
\end{figure*}

Figure~\ref{fig:del_linear} shows a comparison of the sensitivities when
two different energy binning schemes are used (and the $\mu^+$ and $\mu^-$
events were separately binned). It is observed that the choice of linear
energy bins improves the overall sensitivity. This is because the finer
binning at lower energy in the linear bins allowed to better probe the
oscillation signatures.

\subsection{Precision Measurements}
\label{pre_mea}
 
The precision on the oscillation parameters is given by:
\begin{eqnarray}
{\rm Precision}^{n\sigma} = \frac{(P_{max}^{n\sigma} -
P_{min}^{n\sigma})}{(P_{max}^{n\sigma} + P_{min}^{n\sigma})}~,
\end{eqnarray}
where $P_{max}^{n\sigma}$ and $P_{min}^{n\sigma}$ are the maximum
and minimum values of the concerned oscillation parameters at a
given confidence level, $n$. From Tables~\ref{pre1} and ~\ref{pre2}
we conclude that the analysis with charge separation of the muon event
significantly improved the capability of ICAL detector for the estimation
of oscillation parameters. In particular, the $1\sigma$ sensitivity
for $\sin^2\theta_{23}$ ($\Delta m_{32}^2$) improved from 30\% to 27\%
(11.5\% to 10\%) on changing from exponential to linear bins and including
charge identification capability for the muons.

\begin{table*}
  \centering
  \caption{\label{pre1} ICAL's capability and precision reach for measuring
the atmospheric mixing angle $\sin^{2}\theta_{23}$ with a precision of
at $1\sigma$, $2\sigma$ and $3\sigma$ confidence levels respectively
for both the binning schemes.}
\begin{tabular}{|c | c | c|c | c |} \hline 
& \multicolumn{2}{|c|}{Precision with} & \multicolumn{2}{|c|}{Precision with} \\
& \multicolumn{2}{|c|}{exponential bins} & \multicolumn{2}{|c|}{linear
bins} \\
Confidence level & Without cid (\%) & With cid (\%) 
& Without cid (\%) & With cid (\%)  
\\ \hline
$1\sigma$ & 32.0 & 30.0 & 29.5 & 27.0 \\
$2\sigma$ & 46.0 & 43.0 & 43.0 & 39.0 \\
$3\sigma$ & 57.0 & 53.0 & 52.5 & 48.0 \\
\hline
\end{tabular} 

\end{table*}

\begin{table*}
  \centering
  \caption{\label{pre2} ICAL's capability and precision reach for measuring
the atmospheric mass squared difference $\Delta m^{2}_{32}$ with a
precision of at $1\sigma$, $2\sigma$ and $3\sigma$ confidence levels
respectively for both the binning schemes.}
\begin{tabular}{|c | c | c|c | c |} \hline 
& \multicolumn{2}{|c|}{Precision with} & \multicolumn{2}{|c|}{Precision with} \\
& \multicolumn{2}{|c|}{exponential bins} & \multicolumn{2}{|c|}{linear
bins} \\
Confidence level & Without cid (\%) & With cid (\%) 
& Without cid (\%) & With cid (\%)  
\\ \hline
 
$1\sigma$ & 12.7 & 11.5 & 11.3 & 10.0 \\
$2\sigma$ & 26.5 & 25.4 & 24.2 & 23.3 \\
$3\sigma$ & 42.4 & 41.3 & 39.3 & 38.3 \\
\hline 
\end{tabular} 

\end{table*}

\subsection{Allowed region in $\Delta m^2_{32}$-$\sin^2\theta_{23}$
parameter space}

The two dimensional confidence region for the two oscillation parameters
($\Delta m^{2}_{32}$, $\sin^{2}\theta_{23}$) has been determined by
allowing $\Delta m^2_{32}$, $\sin^2\theta_{23}$ and $\sin^2\theta_{13}$
to vary over their $3\sigma$ ranges as shown in Table~\ref{table1}. The
contour plots have been obtained for $\Delta\chi^{2} = \chi^{2}_{min}
+ A$, where $\chi^{2}_{min}$ is the minimum value of $\chi^{2}$ for
each set of oscillation parameters and values of A are taken as 2.30,
4.61 and 9.21 corresponding to 68\%, 90\% and 99\% confidence levels
respectively for two degrees of freedom.

\begin{figure*}[htp]
  \centering
\includegraphics[width=0.49\textwidth]{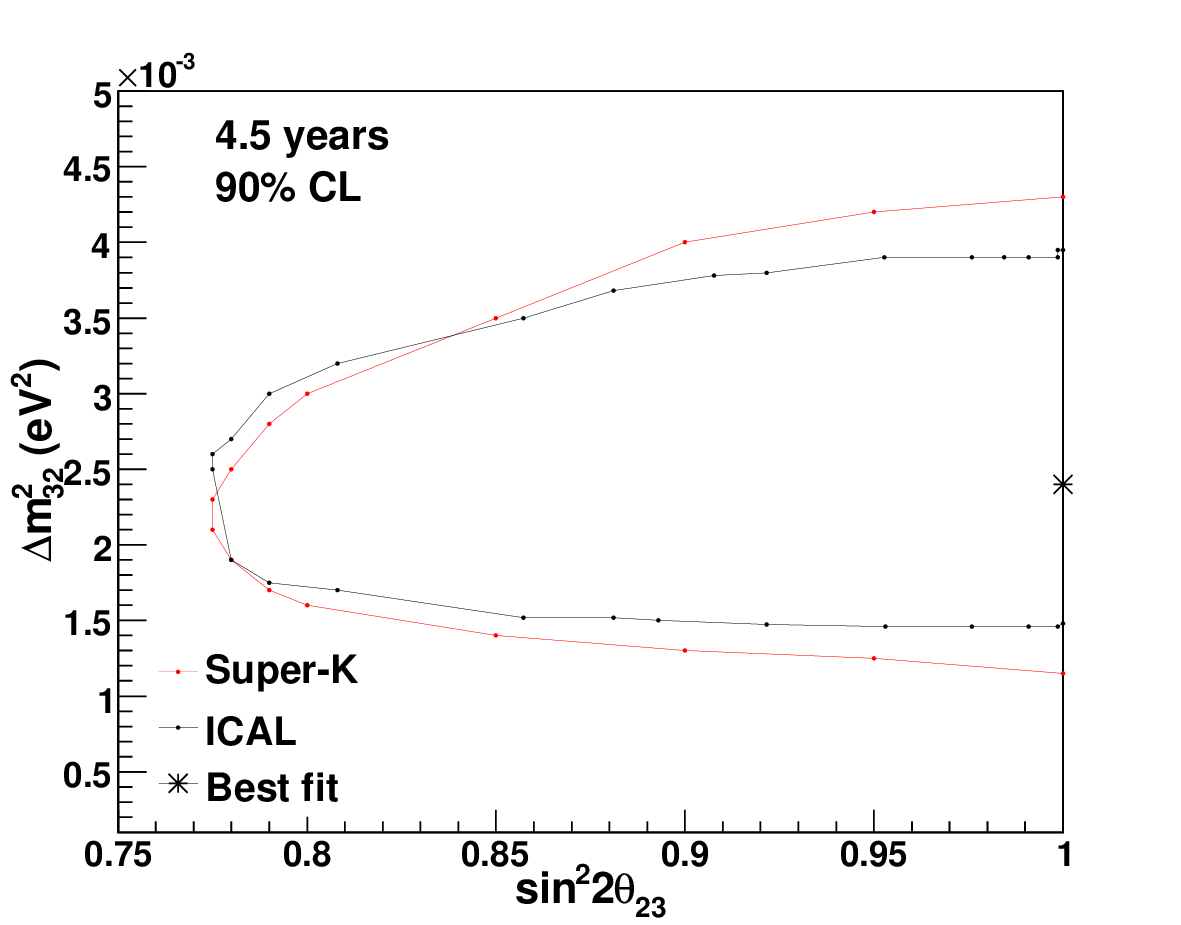}
\includegraphics[width=0.49\textwidth]{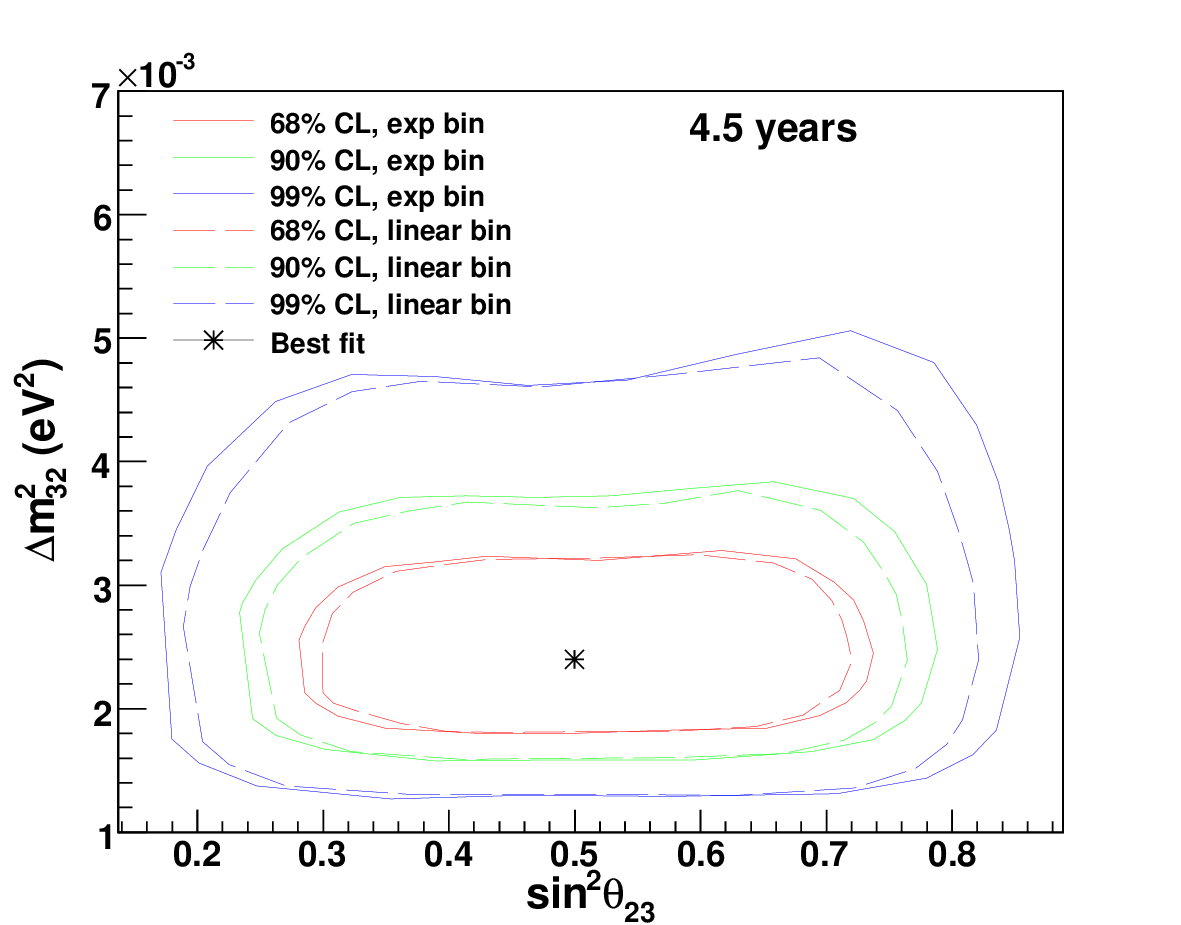}
\caption{The 90 \% CL contour of ICAL for 4.5 years of simulated
data in comparison with Super-Kamiokande data~\cite{nitta}. (left panel). The
precision reach expected at ICAL in the $\sin^{2}\theta_{23}$-$\Delta
m^{2}_{32}$ plane for 4.5 years running of the 51 kton detector using
two different energy binning schemes, viz., exponential and linear,
without charge separation (right panel).}
\label{fig:cont4.5}
\end{figure*}

We have used the systematic uncertainties as described earlier
and the definition of $\chi^{2}$ given in Eq.~\eqref{chi1}. In figure~\ref{fig:cont4.5}, the 90\% CL contour of ICAL for 4.5 years
data simulation (exponential binning scheme and combined $\mu^+$ and
$\mu^-$ events) is compared with Super-Kamiokande data~\cite{nitta}. For
Super-Kamiokande, the 90\% CL allowed region of parameter space is given
by ($\sin^{2}2\theta_{23} \ge 0.765, \Delta m^{2}_{32} = (1.2$--$4.3)
\times 10^{-3}$ eV$^{2}$). For ICAL, the corresponding allowed region is
($\sin^{2}2\theta_{23} \ge 0.771, \Delta m^{2}_{32} = (1.48$--$3.7) \times
10^{-3}$ eV$^{2}$), so ICAL has similar sensitivity as Super-Kamiokande
for the same exposure.

The precision reach expected at ICAL in the $\sin^{2}\theta_{23}$-$\Delta
m^{2}_{32}$ plane for 4.5 years with the exposure of 51 kton
detector using two different energy binning schemes as mentioned in
section~\ref{evt_gen} is also shown in figure~\ref{fig:cont4.5}. Again,
the linear binning scheme is more sensitive than the exponential one.

\begin{figure}[htp]
  \centering
\includegraphics[width=0.5\textwidth]{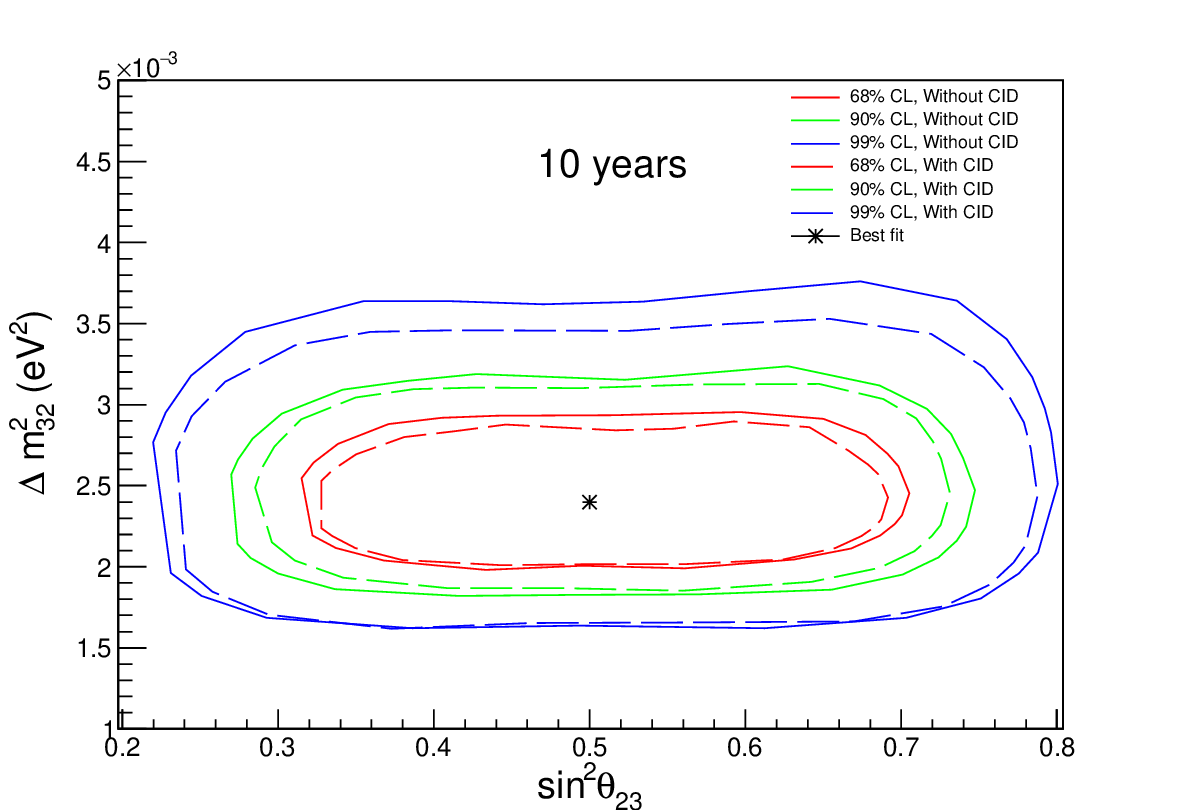}

\caption{The precision reach expected at ICAL in the
$\sin^{2}\theta_{23}$-$\Delta m^{2}_{32}$ plane for
10 years with 51 kton detector, with and without charge identification
for the exponential binning scheme.}
\label{fig:cont5n10-10n11}
\end{figure}

Figure~\ref{fig:cont5n10-10n11} shows the precision reach expected at ICAL
in the $\sin^{2}\theta_{23}$-$\Delta m^{2}_{32}$ plane for 10 years with
the exposure of 51 kton detector, with and without charge identification
for the exponential binning scheme. It is seen that the sensitivity
improves with the addition of charge identification efficiencies in both
$\Delta m^{2}_{32}$ and $\sin^{2}\theta_{23}$.

In summary, it is seen that although the muons lose different amounts
of energy depending on the distance traversed in the rock, these events
are still sensitive to the neutrino oscillation parameters. Due to the
statistical limitations, these events do not have significant sensitivity
to the sign of the 2--3 mass squared difference (sign of $\Delta
m^2_{32}$) or to the octant of $\theta_{23}$ (whether this lies in the
first or second quadrant, or is in fact maximal). Although the sensitivity
is not as significant as that from direct detection of atmospheric
neutrinos in the detector, with such low counting experiments, every
independent source of information needs to be taken into account. Hence
rock muon events provide a useful and independent additional source of
information on the neutrino oscillation parameters in the 2--3 sector.

\begin{figure}[htp]
  \centering
\includegraphics[width=0.5\textwidth]{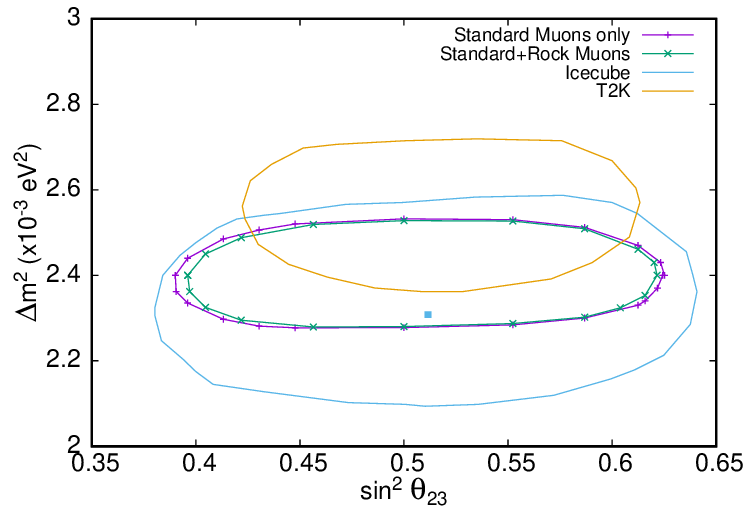}
\caption{Allowed contours at 90\% CL in the $(\sin^{2}\theta_{23}$--$\Delta m^{2}_{32})$ plane for input values of
($\sin^{2}\theta_{23}, \Delta m^{2}_{32}) = (0.5, 2.4\times 10^{-3}$
eV$^2$). The IceCube \cite{icecube} and T2K \cite{t2k}
data at 90\% CL are also shown in comparison.
The blue dot shows the best fit values from
the IceCube data, $(\sin^2\theta_{23},\Delta m^{2}_{32}) = (0.51,
2.31 \times 10^{-3}$ eV$^2)$ with NH.}
\label{fig:cont5n10-10n13}
\end{figure}

The rock muon analysis that we have performed so far uses the events
produced when the atmospheric neutrinos (mostly coming from below)
interact with the rock surrounding the detector. The so-called ``standard
muons'' are those produced when the atmospheric neutrinos directly
interact with the material of the detector and are the main focus of
ICAL. However, as we have shown, the rock muons are also sensitive to
the neutrino oscillation parameters, especially in the 2--3 sector. In
addition, the {\em source} and hence fluxes of neutrinos are the same
in both cases, and the detector response is the same since the primary
detection is through muons in both cases. Hence we can combine the
two sets of events and find their combined sensitivity to neutrino
oscillation parameters. The details of the analysis for standard
muons can be found in Ref.~\cite{lsmohan}. The combined results,
using both standard and rock muon events, are shown in comparison with
IceCube~\cite{icecube} and T2K~\cite{t2k} data (assuming normal ordering)
in figure~\ref{fig:cont5n10-10n13}. Here ICAL data for 10 years has been
considered and it has been assumed for simplicity that the same cross
section systematics applies to both sets. There is a 3--4\% improvement
in the sensitivity to $\sin^2\theta_{23}$ and a barely perceptible
improvement in $\Delta m^2$, the 2--3 oscillation parameters, when
the rock muon information is added to the standard muon analysis. While
the sensitivity of ICAL is comparable to the other experiments, however,
note that both T2K and IceCube are taking data while ICAL is yet to be
built.

\section{Discussions and Conclusion} \label{diss_con}

A Monte Carlo simulation using the NUANCE neutrino generator for 4.5 and
10 years exposure of ICAL detector to upward-going muons, generated by the
interaction of atmospheric neutrinos with the rock material surrounding
the proposed ICAL detector, has been carried out. For this analysis, the
muon momentum and angle resolutions, as well as the reconstruction and
charge identification efficiencies were separately studied using a
GEANT4-based code for a sample of muons entering the bottom part of the
detector, which is relevant for the present study.

The analysis has been done using three neutrino flavor mixing and by
taking Earth matter effects into account;
various selection criteria were also included to reduce the
contribution from the
cosmic ray muon background as well as the
standard charged-current atmospheric muon neutrino
events. A marginalized $\Delta \chi^{2}$ analysis with finer bins at
lower energy has been performed. Various systematic uncertainties have
also been included in the analysis. The ICAL detector results were
compared with Super-K detector for 4.5 years of data and both of them
were comparable. The analysis was done for 10 years of 51 kton exposure
of INO-ICAL detector, with 10 systematic uncertainties, using charge
separation of the upward-going muons. Also, a comparison of the reach of
ICAL with IceCube \cite{icecube} and T2K \cite{t2k} was done and shows a
marginally better sensitivity for ICAL; although it is to be noted that
both Icecube and T2K are already accumulating data.

The main aim of the proposed ICAL detector is to make precision
measurements of neutrino oscillation parameters, especially in the 2--3
sector. Upward-going muons arise from the interactions of atmospheric
neutrinos with the rock material surrounding the detector, and they
carry signatures of oscillation in spite of energy loss of the muon
before they reach the detector. Hence an independent measurement
of the oscillation parameters is provided by upward-going or rock
muons~\cite{upmuon,upmuon1,upmuon2}, although the sensitivity of
upward-going muons to the oscillation parameters is lower than contained
vertex events where the muon neutrinos directly interact with the detector
via charged current interactions to produce muons.

Since the atmospheric neutrino fluxes fall off rapidly with energy
($\sim E^{-2.7}$), studies of conventional contained-vertex events
in ICAL are dominated by low-energy events. In contrast, it is
seen that the upward-going muon sample with a larger proportion of high
energy events have a better probability of reaching the detector. Hence
the contained-vertex and upward-going muons are complementary to each
other. A combined analysis of both sets of events will therefore be useful
to reduce overall errors due to flux and cross section normalization
uncertainties. This is beyond the scope of the current work.

\paragraph{Acknowledgments}:
The authors thank the INO physics group coordinators for their comments
and suggestions on the results and the INO collaboration for their
support and help; also, the referee for very valuable suggestions and
remarks. We also thank S. Pethuraj and Hemalata Nayak for help with cosmic
muon measurements at mini-ICAL and events visualisation. R. Kanishka
acknowledges University Grants Commission, Department of Atomic Energy
and Department of Science \& Technology (Govt. of India) for financial
support.

\end{document}